\documentclass[twocolumn,showpacs,nofootinbib,preprintnumbers,prd]{revtex4-1}

\usepackage{amsmath}
\usepackage{amsfonts}
\usepackage{amssymb}
\usepackage{graphicx}
\usepackage[titletoc]{appendix}
\usepackage{color}
\usepackage{hyperref}
\usepackage{cleveref}
\usepackage[rightcaption]{sidecap}
\usepackage{subfigure}
\usepackage{comment}

\usepackage{mathtools}

\usepackage{dcolumn}

\usepackage{array}
\usepackage{ctable}
\usepackage{multirow}
\usepackage{siunitx}
\usepackage{longtable}
\usepackage{tabularx}
\usepackage{booktabs}
\usepackage{supertabular}
\usepackage{verbatim}

\graphicspath{{Graphics/}}

\def\be{\begin{equation}}
\def\ee{\end{equation}}
\def\bea{\begin{eqnarray}}
\def\eea{\end{eqnarray}}
\renewcommand{\d}{\mathrm{d}}

\definecolor{vividviolet}{rgb}{0.62, 0.0, 1.0}
\definecolor{amaranth}{rgb}{0.9, 0.17, 0.31}
\definecolor{palatinateblue}{rgb}{0.15, 0.23, 0.89}
\definecolor{brightpink}{rgb}{1.0, 0.0, 0.5}
\definecolor{cornflowerblue}{rgb}{0.39, 0.58, 0.93}
\definecolor{deepcarminepink}{rgb}{0.94, 0.19, 0.22}
\definecolor{radicalred}{rgb}{1.0, 0.21, 0.37}

\hypersetup{ linktoc=all,
    colorlinks, linkcolor={palatinateblue},
    citecolor={brightpink}, urlcolor={amaranth}
}

\begin{document}

\title{Gravitational metamaterials from optical properties of spacetime media}

\author{Orlando Luongo}
\email{orlando.luongo@unicam.it}
\affiliation{Universit\`a di Camerino, Via Madonna delle Carceri, Camerino, 62032, Italy.}
\affiliation{Department of Nanoscale Science and Engineering, University at Albany-SUNY, Albany, New York 12222, USA.}
\affiliation{Istituto Nazionale di Astrofisica (INAF), Osservatorio Astronomico di Brera, 20121 Milano, Italy.}
\affiliation{Istituto Nazionale di Fisica Nucleare (INFN), Sezione di Perugia, Perugia, 06123, Italy,}
\affiliation{Al-Farabi Kazakh National University, Al-Farabi av. 71, 050040 Almaty, Kazakhstan.}

\begin{abstract}
Gravitational optical properties are here investigated under the hypothesis of spherically-symmetric spacetimes behaving as media. To do so, we first consider two different definitions of the refractive index, $n_O$, of a spacetime medium and show how to pass from one definition to another by means of a coordinate transformation. Accordingly, the corresponding physical role of $n_O$ is discussed by virtue of the Misner-Sharp mass and the redshift definition. Afterwards, we discuss the inclusion of the electromagnetic fields and the equivalence with nonlinear effects induced by geometry. Accordingly, the infrared and ultraviolet gravity regimes are thus discussed, obtaining bounds from the Solar System, neutron stars and white dwarfs, respectively. To do so, we also investigate the Snell's law and propose how to possibly distinguish regular solutions from black holes. As a consequence of our recipe, we speculate on the existence of \emph{gravitational metamaterials}, whose refractive index may be negative and explore the corresponding physical implications, remarking that $n_O<0$ may lead to invisible optical properties, as light is bent in the opposite direction compared to what occurs in ordinary cases. Further, we conjecture that gravitational metamaterials exhibit a particle-like behavior, contributing to dark matter and propose three toy models, highlighting possible advantages and limitations of their use. Finally, we suggest that such particle-like configurations can be ``dressed" by interaction, giving rise to \emph{geometric quasiparticles}. We thus construct modifications of the quantum propagator as due to nonminimal couplings between curvature and external matter-like fields, finding the corresponding effective mass through a boson mixing mechanism.
\end{abstract}


\maketitle
\tableofcontents

\section{Introduction}

The optical properties induced by gravitational fields have been crucial in checking the goodness of general relativity since the Eddington pioneering experiment more than one century ago  \cite{Wambsganss:1998gg}.

Currently, studying the optical properties of spacetime is particularly important for identifying potential deviations from Einstein's theory \cite{Will:2014kxa,Capozziello:2019cav}. These effects can be detected within the regime of strong gravity\footnote{At present, there \emph{is not a net evidence} in favor of extending gravity, albeit various attempts, spanning from $f(R)$ \cite{DeFelice:2010aj,Sotiriou:2008rp}, $f(T)$ \cite{Bahamonde:2021gfp}, $f(Q)$ \cite{Heisenberg:2023lru}, up to experimental analyses \cite{Aviles:2012ay,Calza:2019egu,Capozziello:2014zda,Capozziello:2015rda}, phase-space dynamics \cite{Carloni:2024rrk,Carloni:2023egi,Paliathanasis:2023nkb,Dimakis:2022rkd,Khyllep:2021pcu,Paliathanasis:2016vsw}, and so on, have been proposed; see e.g. \cite{Bamba:2012cp}. A remarkable indirect exception is offered by the Planck measurements that have suggested as best-suited inflationary potential the Starobinsky potential \cite{Planck:2018jri}, conformally equivalent to a second order extension of gravity; see, e.g. \cite{Belfiglio:2024swy,Luongo:2023aaq}. }, using techniques of gravitational optics and/or lensing, seeking possible clues against general relativity.

Naively, the first attempt to tackle gravitational optical properties is to explore light propagation in the Schwarzschild metric \cite{optical}. Accordingly, the so-called Regge-Wheeler equation has provided valuable solutions both above and below black hole event horizons \cite{Boonserm:2013dua,Fiziev:2007es}, \textit{i.e.}, uncovering that the region below the event horizon could be filled with electromagnetic disturbances, providing \emph{de facto} light rays trying to cross the horizon from below \cite{ValienteKroon:2007bj}.

Even though the physical significance of the Regge-Wheeler treatment is well-consolidate, debated alternatives, that bypass the direct inclusion of the electromagnetic field, have successfully been used to \emph{directly} characterize the light rays around gravitational sources \cite{Wheeler:1955zz}. The basic demands of any alternative proposal consist in ensuring standard thermodynamics to hold \cite{Jacobson:1995ab} and in imaging that spacetimes behave similarly to materials under precise circumstances \cite{Turimov:2023man}.

In particular, focusing on the cited black holes, they are currently reviving novel great enhancements in view of the recent observations of black hole shadows \cite{Bronzwaer:2021lzo}. In general, black holes are also associated with naked singularity counterparts, arising when event horizons do not  form. The Schwarzschild spacetime with negative mass is the simplest example and the prototype of such configurations. The cosmic censorship conjecture \cite{Penrose:1969pc}, however, limits their existence, suggesting they may be purely mathematical constructs. Conversely, alternatives suggest they could emerge or influence gravitational collapse and, even, contribute to the observed cosmic speed up, inducing repulsive gravity \cite{Luongo:2010we,Luongo:2015zaa}.

Hence, again, the interest in theories that can extend general relativity, violating, for example, the Penrose collapse theorems have acquired renewed interest in the most recent literature.

In this perspective, regular black holes can challenge the standard picture of black hole formation \cite{Kerr:2023rpn,Malafarina:2022wmx}, as they may emerge from nonlinear electrodynamics \cite{Zaslavskii:2010qz,1966JETP...22..241S,1966JETP...22..378G}, albeit remarkable examples suggest their pure gravitational origin \cite{Bueno:2024dgm}. In addition, the debate about their existence is currently ongoing, with promising indirect results in framing observational effects, such as quasi-periodic oscillations \cite{Boshkayev:2023rhr,Boshkayev:2022haj}, quasi-normal modes \cite{Berti:2009kk}, mimickers \cite{Casadio:2024lgw}, etc.

Nevertheless, astrophysical black holes are usually described either by Schwarzschild or Kerr metrics, following the so-called \emph{Kerr hypothesis}. However, this lacks strong experimental confirmation \cite{Cardoso:2016ryw,Yagi:2016jml} and can be firmly established once precise observations of the gravitational fields around black hole candidates are rigorously tested against predictions derived from these spacetimes.

Following this conceptual point, as stated above, black hole gravitational fields can be modeled as media and, in analogy, it is also possible to model materials through spacetime solutions \cite{Giddings:2002kt,Bleicher:2011uj}. In this respect, recent advances in theoretical and experimental physics have indicated \emph{metamaterials} as a frontier for exotic new materials \cite{2024RvMP...96a5002Y}, which are able to exhibit unusual optical properties that can, in principle, even be reproduced from spacetime metrics.

On the other side, cosmology is frustratingly experiencing conceptual and experimental controversies that jeopardize the standard background model, suggesting possible new physics behind cosmological tensions \cite{Cortes:2023dij,Abdalla:2022yfr,DiValentino:2021izs}, dark energy \cite{Copeland:2006wr,Wolf:2023uno,Peebles:2002gy}, dark matter \cite{Bertone:2018krk,Cirelli:2024ssz,Lisanti:2016jxe}, and so on. Thus, despite decades of intense speculation, a definitive solution seems no closer to explain what the dark universe is made of\footnote{ Conflicting measurements are currently in favor of a possible slightly evolving dark energy term \cite{Carloni:2024zpl,Luongo:2024fww,Alfano:2024jqn}, although in net tension with the use of $\Lambda$, and hints for ultralight dark matter candidates \cite{Ferreira:2020fam}, that may partially exclude weakly interacting massive particles}.

Motivated by the above considerations, in this paper we investigate optical properties and the effective refractive index $n_O$, of gravitational sources, modeled by black holes, regular black holes, naked singularities and solutions with no horizons, adopting two different methods. We first obtain the refractive index using direct ray propagation, whereas the second is obtained by employing an optical metric, where the refractive index appears as an unknown function. Once demonstrated the equivalence between the two approaches, we explore the refractive index definition to black holes, namely to the Schwarzschild, Reissner-Nordstr$\ddot o$m and Schwarzchild-de Sitter solutions. Analogously, we compute the refractive indexes for regular black holes, namely working out the Bardeen, Hayward, Dymnikova and Fang-Wang solutions. We emphasize the main differences among the various cases, comparing our findings with the Schwarzschild case and trying to remark how to optically distinguish black holes from mimickers. Considering spacetimes as media, we then investigate the Snell's law for standard stars, invoking our Sun as benchmark. Later, the case of strong gravitational sources is also studied, employing neutron stars and white dwarfs. In so doing, we find constraints and upper bounds on the free parameters of the underlying metrics, guaranteeing the viability of Snell's law and refractive index definitions.

Accordingly, we explore the case of negative refractive indexes inspired by the well-known classes of exotic materials, dubbed \emph{metamaterials}, assuming that spacetimes may behave similarly. We thus baptize the spacetimes that exhibit negative refraction \emph{gravitational metamaterials} to briefly indicate those metrics that mime the properties of metamaterials.

A subsequent physical interpretation of this landscape is discussed, including the role played by the Misner-Sharp (MS) mass and the redshift definition. In this respect, we speculate about the theoretical viability of having a negative refractive index from a spacetime medium, without altering the Snell's law.

The optical representation of $n_O$ is thus compared with the expectations of electromagnetic fields in curved space. The direct analogy between the two approaches, i.e., ours involving general relativity, and the other assuming the electric and magnetic fields, is remarked, showing under which conditions the two landscapes coincide. Last but not least, the homogeneous, inhomogenous and anisotropic contributions to the refractive indexes are thus examined, corresponding to constant, i.e., independent from the $r$ coordinate, and $r$-dependent functions. In addition, we analyze the same outcomes in flat space, involving as a remarkable example the Rindler coordinates. Even in flat space, we find hints on the possibility of having negative refraction, defining a viable Unruh temperature.

Afterwards, we speculate toward the particle-like nature of those objects that fulfill the Snell's law but violates the positivity of $n_O$. Accordingly, we physically propose that gravitational metamaterials can exist under the form of invisible compact objects, if they turn out to be regular, horizon-free and whose energy-momentum tensor can be constructed to mimic a particle. We pursue this target in comparison with black holes under the form of particles, as proposed in Ref. \cite{Holzhey:1991bx,Arkani-Hamed:2019ymq}. In the picture in which energy is picked and pressure is negligible, we speculate that dark matter may be composed of \emph{no-horizon regular solutions, with negative refractive index, behaving as gravitational metamaterials}. Three main examples are thus discussed, as possible toy models to characterize the particle-like natures of such solutions. The first involves an homogeneous refractive index, the second the Newtonian metric as a remarkable hint of inhomogeneous $n_O$, while the third case is intimately connected to the construction of no-horizon regular solutions, involving the Simpson-Visser procedure of regularization \cite{Simpson:2018tsi}.

Finally, we seek the quantum nature of these particle-like settings and conjecture how to extend them in terms of \emph{quasiparticles of geometry}. Hence, involving two matter-like scalar fields, with vanishing sound speeds to emulate dust, we compute the boson mixing and the new states corresponding to the quasiparticle configurations. Theoretical consequences of our recipe and observable predictions are then reported across the text.

The paper is organized as follows. In Sect. \ref{sezione2}, we formulate the refractive index definition for spherically symmetric metrics, with the hypothesis that spacetime behaves like a medium. We review the $n_O$ definitions in view of the MS mass, redshift, and  weak field regime. The conditions to provide negative refractive indexes are also explored. In Sect. \ref{sezione3}, we incorporate the electromagnetic fields and introduce the concept of anisotropic refractive index. Afterwards, the optical properties of black holes, regular black holes and naked singularities are investigated in Sect. \ref{sezione4}. Further, in sect. \ref{sezione5}, we work out the Snell's law to infrared and ultraviolet regimes of gravity, respectively for the Sun, neutron stars and white dwarfs. The role of negative refraction is also studied in flat spaces, employing the Rindler coordinates in Sect. \ref{sezione6}, moreover defining a viable definition of the Unruh temperature. Last but not least, we conjecture the existence of stable particle-like configurations in Sect. \ref{sezione7}. Based on it, in Sect. \ref{sezione8}, three toy models are, thus, proposed and checked. There, speculating that gravitational metamaterials may contribute to dark matter, we thus seek their quantum nature and propose how they turn out to be quasiparticles of geometry. Finally, Sect. \ref{sezione9} is devoted to conclusions and perspectives.

\section{Optical properties of  spherical spacetimes}\label{sezione2}

The optical properties of a generic spacetime can be explored by investigating how spacetime acts as a medium \cite{Celoria:2017idi, Celoria:2017bbh, Ballesteros:2016kdx, Ballesteros:2016gwc}. In analogy to classical physics, the optical properties of a medium can be described even without directly including the electromagnetic properties of light. This is the prerogative of \emph{geometrical optics} that does not consider the electromagnetic Lagrangian but rather the consequences on different media, when the light beam is associated with a particle rather than a wave \cite{book}.

Bearing this in mind, the corresponding refractive index, $n_O$, turns out to be positive and larger than unity. Throughout this paper, we will explore the possibility that \emph{$n_O$ is also negative and we will discuss some of the corresponding properties}.

To investigate $n_O$, we consider two treatments able to extract the optical properties of a spacetime medium once the metric is fixed. We then compare the results obtained from each approach and show that they turn out to be equivalent, up to a given coordinate transformation.

Later, we restore the use of electromagnetic field, remarking that it is possible to mime the electromagnetic effects through general relativity alone.

In particular, we consider the two scenarios below.

\begin{itemize}

\item[-] The first treatment to obtain optical properties ensures the right behavior of the motion of light beams, without passing through the electromagnetic field \cite{optical}. Consequently, the optical refractive index can be directly inferred. For this reason, the method is here conventionally referred to as the \emph{direct approach}.

\item[-] The second strategy aims to parameterize the medium by presuming \emph{a priori} the existence of a refractive index, incorporated as a function of the metric itself. The method is here conventionally dubbed  \emph{indirect approach}.

\end{itemize}

For each method, we can then
\begin{itemize}
    \item[-] show under which circumstances the spacetime can be modeled in terms of a medium and,
    \item[-] describe how the spacetime can resemble the properties of exotic materials.
\end{itemize}
The second point consists of justifying whether the refractive index can become negative, \textit{i.e.}, a prerogative of the so-called \emph{metamaterials} \cite{revmeta}.

We begin by summarizing the two strategies below, and then discuss each in detail for specific spacetimes, whose metric components are given by  $g\equiv diag(g_{00},g_{11},g_{22},g_{33})$.

\subsection{Direct strategy}

In the direct method, we assume that light propagates in a medium that is mimicked through a given spacetime.

Consider two coordinate systems, $(T,L,\theta,\phi)$ and $(t,l,\theta,\phi)$. Thus, we can write
\begin{subequations}
\begin{align}
&ds^2=dL^2-c^2dT^2\,,\label{prototype}\\
&ds^2=dl^2-c^2dt^2\,,\label{prototype2}
\end{align}
\end{subequations}
and, so, a light beam, corresponding to $ds=0$, is recovered in both the two spacetimes, giving immediately $\frac{d L}{dT}=\frac{dl}{dt}=c$.

Now, consider, in the simplest static spherically symmetric configuration, the transformations
\begin{subequations}
    \begin{align}
&dL^2=\beta^2dr^2+\gamma^2r^2d\Omega^2,\label{2a}\\
&dT^2=c^2\alpha^2dt^2,\label{2b}
    \end{align}
\end{subequations}
where $\alpha=\alpha(r),\beta=\beta(r)$ and $\gamma=\gamma(r)$ are unspecified functions. As usual, we only require that they cannot change sign to prevent them from vanishing. Moreover, \emph{a priori}, we do not assume that $\alpha$, $\beta$, and $\gamma$ are positive definite, leaving open the possibility that they may be negative, as we will clarify later.

It is therefore possible to argue a relation between the two sets of coordinates, namely it is possible to write the coordinates $(T,L,\theta,\phi)$ in terms of  $(t,l,\theta,\phi)$.

Nevertheless, from Eq. \eqref{prototype}, with  Eqs. \eqref{2a}-\eqref{2b}, we end up with
\begin{equation}\label{metricagenerica}
ds^2=\beta^2dr^2+\gamma^2r^2d\Omega^2-c^2\alpha^2dt^2\,,
\end{equation}
where  $d\Omega^2\equiv d\theta^2+\sin^2\theta d\phi^2$.

Hence, assuming that a light beam is moving in the gravitational field induced by the above metric and considering a relation to relate the coordinates $(L,T)$ to $(l,t)$, we can easily parameterize $dr$ and $rd\Omega$ by introducing a \emph{deflection angle}, $\sigma$, as
\begin{subequations}\label{parametric}
\begin{align}
dr&=\cos\sigma dl,\\
rd\Omega&=\sin\sigma dl,
\end{align}
\end{subequations}
that is the standard procedure used in classical physics, when dealing with geometrical optics.

Easily, we obtain
\begin{subequations}
    \begin{align}
        dL^2&\equiv \beta^2\cos^2\sigma dl^2+\sin^2\sigma\gamma^2 dl^2,\label{parte1}\\
        dT^2&\equiv \alpha^2dt^2.\label{parte2}
    \end{align}
\end{subequations}

\noindent Now, divide Eq. \eqref{parte1} with  Eq. \eqref{parte2},
\begin{equation}\label{definizionen}
\frac{dL}{dT}=\frac{\beta}{\alpha}\sqrt{1-\sin^2\sigma\left(1-\frac{\gamma^2}{\beta^2}\right)}\frac{d l}{d t},
\end{equation}
defining consequently an effective representation of the refractive index provided by
\begin{equation}\label{indicedirifrazione1}
n_O\equiv\frac{dL}{dT}\left(\frac{dl}{dt}\right)^{-1}=\frac{c}{v}=\frac{\beta}{\alpha}\sqrt{1-\sin^2\sigma\left(1-\frac{\gamma^2}{\beta^2}\right)},
\end{equation}
where $n_O$ turns out to be the \emph{gravitational refractive index}, found by following the standard recipe to provide  the ratio between the speed of light and $v$, \textit{i.e.}, the velocity seen by the second observer \cite{optical}.

Immediately, fixing $\beta$ as positive, the sign of $\alpha$ discriminates whether $n_O>0$ or $n_O<0$, as we will discuss later in the text.

\subsection{Indirect strategy}

The indirect strategy assumes \emph{a priori} the existence of a refractive index inside the metric.

In particular, it is possible to reformulate the Fermat’s principle of shortest optical paths, ensuring that the path taken by a ray between two spacetime events is the path that can be traveled in the least time.

Assuming its validity for geodesic trajectories of given spacetimes, it is possible to obtain an equivalent  description of
light rays in a gravitational field than classical
optics.

Introducing, then, the isotropic coordinates, $(t,\hat r,\theta,\phi)$, namely
\begin{equation}
\hat r=r \exp \left\{ \int_r^\infty \left(1-\left| \frac{\beta}{\gamma }\right| \right)
\frac{d r'}{r'} \right\},
\label{eq:r-rho}
\end{equation}
with $\gamma^2(\infty)=\beta^2(\infty)$, allowing one to rewrite Eq. \eqref{metricagenerica} as
\begin{eqnarray}\label{metricagenerica2conforme}
d s^2=-g_{00}[r(\hat r)] c^2\, d t^2+\Lambda(\hat r) \, (d\hat r^2 +
\hat r^2\Omega^2),
\end{eqnarray}
providing the
\emph{isotropic refractive index}. We can thus project the metric into a flat
background and, then, for fixed $\theta$ and $\phi$, the isotropic refractive index becomes
\begin{equation}\label{indiceotticoesplicito}
c^2\left(\frac{d \hat r}{d t}\right)^{-2}=\frac{c^2}{v^2}=\frac{\Lambda}{g_{00}},
\end{equation}
having used the fact that the speed of light is seen with velocity $v$ by the observer in $(t,\hat r,\theta,\phi)$.

To reproduce the same result, reported in Eq. \eqref{indicedirifrazione1}, we rewrite
\begin{subequations}
    \begin{align}
        &\Lambda=f^2n_O^2,\label{lam}\\
        &g_{00}=f^2,\label{f2}
    \end{align}
\end{subequations}
where $f=f(\hat r)$ and $n_O=n_O(\hat r)$  are unknown function, written this time in function of $\hat r$, instead of $r$. Consequently, we rewrite Eq. \eqref{metricagenerica2conforme} as
\begin{equation}\label{metricaottica}
ds^2=f^2(\hat r)n_O^2(\hat r)(d\hat r^2+\hat r^2d\Omega^2)-c^2f^2(\hat r)dt^2,
\end{equation}
and so, comparing the initial spacetime, Eq. \eqref{metricagenerica}, written in coordinates $(t,r,\theta,\phi)$, with Eq. \eqref{metricaottica}, we write
\begin{subequations}
    \begin{align}
    f(\hat{r})&=\alpha(r)\,, \label{condtime}
    \\
    f(\hat{r})n_O(\hat{r})d\hat{r}&=\beta(r) dr\,, \label{condlong}
    \\
    f(\hat{r})n_O(\hat{r})\hat{r}&=\gamma(r)\,. \label{condangular}
    \end{align}
\end{subequations}

The above set can be solved into two steps,

\begin{itemize}
    \item[I.] by dividing Eq.~\eqref{condlong} by Eq.~\eqref{condangular}, obtaining a differential equation, say $\frac{d\hat r}{\hat r}=\frac{\beta}{\gamma}dr$ that, once solved, allows us to write $\hat{r}=\hat{r}(r)$, fixing an integration constant, namely $\hat r_0$;
    \item[II.] having the constraint in Eq. \eqref{condtime}, substituting the above results in Eq. (\ref{condangular}), we write  \cite{Capozziello:2022ygp}
\begin{equation}\label{opticalbis}
    n_O=n_{0}\frac{\gamma(r)}{\alpha(r)}\exp{\left\{-\int \frac{\beta}{\gamma}dr\right\}},
\end{equation}
\end{itemize}
with $n_{0}\equiv \hat r_0^{-1}$.

Hence, for each metric of the direct method, there exists a direct correspondence with the optical spacetime, reported in Eq. \eqref{metricaottica}.

Immediately, fixing $\gamma$ as positive, the sign of $\alpha$ discriminates whether $n_O>0$ or $n_O<0$, as we will discuss later in the text.

\subsection{Negative refractive index}

The formal equivalence between the two methods confirms the validity of the direct approach in reproducing the results of the optical metric, up to a coordinate transformation. Given the simpler mathematical structure of Eq. \eqref{indicedirifrazione1} with respect to Eq. \eqref{opticalbis}, it is therefore convenient to adopt the first definition,  henceforth.

\begin{figure}
    \centering
    \includegraphics[scale=0.4
]{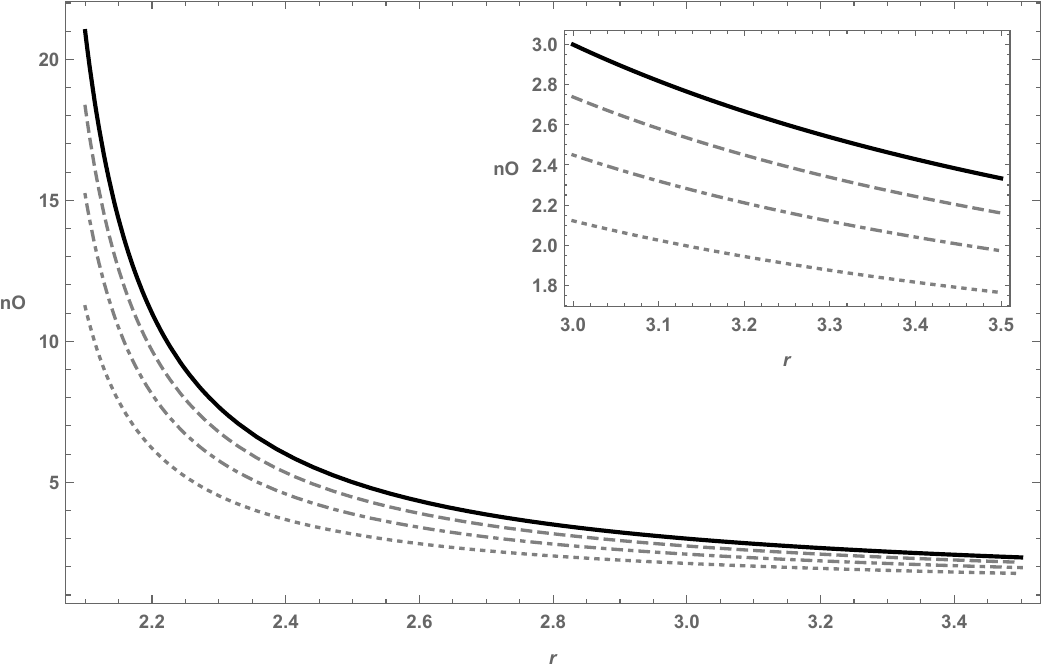}
    \includegraphics[scale=0.45
]{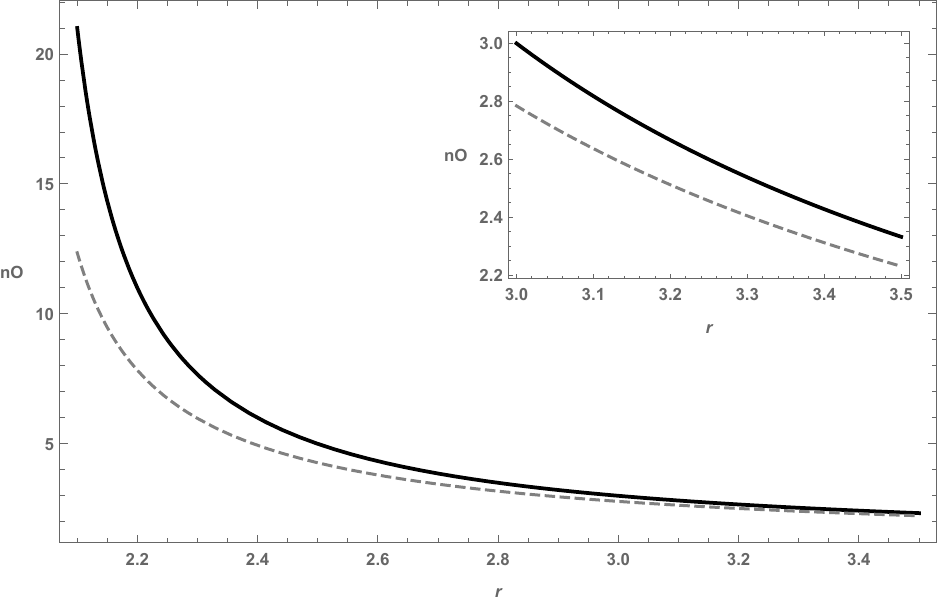}
    \caption{(Top.) Plots of Eqs. \eqref{nS}, varying the angle $\sigma$. The thick curve corresponds to the case $\sigma=0$, namely a radial approximation. The dashed, dotdashed and dotted curves are respectively displayed with $\sigma=\frac{\pi}{3}$, $\sigma=\frac{\pi}{4}$ and   $\sigma=\frac{\pi}{6}$, all with unitary mass. The output shows that only the amplitude of $n_O$ is influenced by varying $\sigma$, while the shape remains unaltered. (Bottom.) Plots of the two definitions of refractive indexes, Eqs. \eqref{indicedirifrazione1} and \eqref{opticalbis}, corresponding to the thick and dashed lines, respectively. The two shapes are essentially indistinguishable, however with different magnitudes. }
    \label{indicialvariareditau}
\end{figure}

Hence, from Eq. \eqref{indicedirifrazione1}, it is possible to state that
\begin{itemize}
    \item[-] regardless of the explicit forms of $\alpha, \beta$ and $\gamma$, the optical index can be positive or negative, \textit{i.e.}, there is no limitation \emph{a priori} that forces $n_O$ to be positive, as in classical physics. Accordingly, $n_O$ is positive (negative) once the ratio ${\beta\over \alpha}$ is positive (negative);
    \item[-] in analogy to geometric optics, if $n_O$ is negative, to avoid superluminal occurrences, it is necessary that $n\leq-1$, since $c\geq |v|$. However, while  dealing with $n_O$ defined from the free spacetime functions, it is not possible \emph{a priori} to discard the case $|n_O|<1$. Moreover, the case $|n_O|=1$ \emph{does not} forcedly corresponds here to the Minkowski metric \emph{only}. Indeed, we have
    \begin{equation}
\beta^2-\alpha^2=\sin^2\sigma(\beta^2-\gamma^2),
\end{equation}
where, as expected,  $\alpha=\beta=\gamma=1$, \textit{i.e.}, the Minkowski metric, turns out to be a particular case. In general, however, if $\beta=\alpha\Rightarrow$ $\gamma=\beta$, with $\sigma\neq0$, leading to a \emph{conformal metric}
\begin{equation}\label{primaconforme}
    ds^2=\alpha^2(-c^2dt^2+dr^2+r^2d\Omega^2),
\end{equation}
providing an equation of state well-known in cosmic strings \cite{Hindmarsh:1994re}.

Conversely, if $\sigma=0$ then $\gamma\neq\beta$, and vice versa and, moreover, since $|\sin\sigma|\leq1$, if $\gamma\rightarrow\alpha$, then $\sigma\rightarrow\frac{\pi}{2}$.

\item[-] Further, the refractive index depends upon the angle $\sigma$. This angle, however, represents a parametric decomposition of coordinates, as reported in Eqs. \eqref{parametric}. Consequently, its arbitrariness just changes the shape of $n_O$. An example of this fact is displayed in Fig. \ref{indicialvariareditau}, where we plot $n_O$ in the Schwarzschild case, with different values of $\sigma$. Interestingly, let us consider the \emph{limiting case}  $\sigma\rightarrow \frac{\pi}{2}$, thus having,
\begin{equation}
    n_O=\frac{\beta}{|\beta|}\frac{|\gamma|}{\alpha }=\pm\frac{|\gamma|}{\alpha},
\end{equation}
that, compared with Eq. \eqref{opticalbis}, with positive $\beta$, indicates that,
\begin{align}\label{condizionigammaalfa}
    n_O>0,\quad {\rm if}\, \gamma>0\,\,\,\, {\rm and}\,\,\,\, \alpha>0,\\
    n_O<0,\quad {\rm if}\, \gamma>0\,\,\,\, {\rm and}\,\,\,\, \alpha<0.
\end{align}
Nevertheless, the case $\gamma<0$ is less interesting since $\gamma$ multiplies the positive 3-sphere volume associated with the metric. Thus, $n_O$ can be therefore negative if the sign of $\alpha$ and $\beta$ are opposite, \textit{i.e.}, ensuring $\alpha<0$ leads to $\beta>0$ for having a negative refraction.
\end{itemize}

\subsection{Inhomogeneous and anisotropic refractive indexes}

Quite interestingly,  Eq. \eqref{indicedirifrazione1} can be recast in the form, $n_O\equiv\frac{\beta}{\alpha}\mathcal F(\sigma,\beta,\gamma)$, where $\mathcal F$ is functional independent of $\alpha$. In cases of particular interest, for example in gravitational lensing, one easily finds
\begin{align}\label{smallconditions}
        \sigma\ll1,\quad \frac{\gamma^2}{\beta^2}\simeq 1,
\end{align}
where the first condition, over the angle, is typical for deflection around strongly massive objects, while the second depends on the spacetime under exam.

Indeed, in the simplest Schwarzschild gravitational field, the conditions in Eq. \eqref{smallconditions} can be easily checked. There, $\gamma=1$ and $\beta^2=\left(1-\frac{2M}{c^2r}\right)^{-1}\simeq 1+\frac{2M}{c^2r}$. So, far enough from the horizon, $r_H\equiv 2M/c^2$, say $r\gg r_H$, the Schwarzschild metric guarantees Eq. \eqref{smallconditions}.

In view of these considerations, it could be useful to single out two different parts of $n_O$, namely an \emph{inhomogenous refractive index}, $n^{inh}_O$, and a  \emph{anisotropic refractive index}, $n^{anis}$. To do so, imposing $\sigma\ll1$, we write,
\begin{align}
    n_O^{inh}&=\frac{\beta}{\alpha}\equiv\sqrt{\frac{g_{11}}{g_{00}}},\label{inh}\\
    n_O^{anis}&=-\frac{\sigma^2}{2}n_O^{inh}\left(1-\frac{\gamma^2}{\beta^2}\right),\label{anis}
\end{align}

Here, the concepts of inhomogeneous and anisotropic contributions to $n_O$ \emph{have nothing to do with the original metric} in Eq. \eqref{metricagenerica}. Inhomogeneous implies  $n^{inh}=n^{inh}(r)$, namely $n_O$ is not a constant as it happens for homogeneous media or perfect lens, while the name ``anisotropic part'' will be clarified later, while incorporating the electromagnetic fields. Remarkably, $n^{anis}$ is associated with the angular part of the metric, as clear from the presence of $\gamma$. So, the introduction of an electromagnetic field will give rise to a matrix whose off-diagonal components correspond to the terms now proportional to $\gamma$.

Hence, by construction, the prominent contribution derives from the inhomogenous part, Eq. \eqref{inh}, that suggests, moreover, that a \emph{homogeneous refractive index} can be obtained as $\beta=\mu\alpha$, with $\mu$ a generic real constant\footnote{In the previous sections, we adopted the units $G=1$. Henceforth, we also assume $G=c=1$ for the sake of simplicity, unless otherwise indicated.}.

For the sake of completeness, it is not possible to neglect the second term \emph{a priori} for \emph{every spacetime}. Indeed, there could exist a spacetime that has
\begin{equation}
n_O\simeq \pm\left|\frac{\gamma}{\alpha}\right|\sin\sigma,\quad {\rm if\,\,\,} \frac{\gamma}{\beta}\gg 1,\quad {\rm with}\,\,\, \sigma\rightarrow\frac{\pi}{2},
\end{equation}
leading to $
    n_O<0$ with $\alpha<0$,
implying that the refractive index can be negative \emph{either if it absorbs all the light traveling or if it bends in the other direction the light itself}, in analogy to solid state physics\footnote{Usually, in laboratory settings, absorption is typically associated with the imaginary part (positive or negative) of the refractive index. For negative refraction, the energy flow, described by the Poynting vector, points in the opposite direction to the phase velocity. Consequently, light is generally bent in the opposite direction compared with what occurs in ordinary materials.}.

We will show later that to require a direct analogy between the electromagnetic field and gravity can be obtained assuming that the \emph{anisotropic} contribution is smaller than the \emph{inhomogeneous} part.

\subsection{Refractive index and MS mass}

We here explore possible physical explanations of our refractive indexes, positive or negative, in the context of general relativity alone, \textit{i.e.}, without including the electromagnetic fields, that will be studied later in the text.

To construct viable refractive indexes, consider
\begin{subequations}\label{SCHW}
    \begin{align}
&\alpha=\left(1-\frac{2M}{r}\right)^{\frac{1}{2}},\\
&\beta=\alpha^{-1},\\
&\gamma=1,
    \end{align}
\end{subequations}
for example in the simplest case of the  Schwarzschild solution. We thus find from Eqs. \eqref{indicedirifrazione1} and \eqref{opticalbis}, respectively,
\begin{subequations}\label{nStot}
    \begin{align}
       n_O^{D}&=\frac{r \sqrt{1 - \frac{2 M \sin^2\sigma}{r}}}{r-2 M},\label{nS}\\
       n_O^{I}&=\frac{\bar n_0re^{-2 \tanh^{-1}\left(\sqrt{1 - 2 M/r}\right)}}{\sqrt{1 - 2M/r}},\label{nSbis}
    \end{align}
\end{subequations}
where the superscripts stand for \emph{direct and indirect methods}. Here, we need to impose $\bar n_0\equiv\frac{2}{M}$, in order to guarantee that at $r\rightarrow\infty$, $n_O^{I}\rightarrow1$.

This choice ensures that at $r\gg r_H$ the metric becomes flat and, therefore, $n_O\simeq \beta/\alpha\simeq 1$. For our purposes, it is naively easier to handle the definition \eqref{indicedirifrazione1} and the case $\sigma=0$, rather the indirect definition, as already stated above.

Consider now the MS definition for Eq. \eqref{metricagenerica}
\begin{equation}\label{MSdefinition}
    m_{MS}(r)=\frac{\gamma r}{2}\Big[1-\beta^{-2}\left(\frac{d}{dr}(r\gamma(r))\right)^2\Big].
\end{equation}

Hence, inspired from the Schwarzschild example, one can wonder how to connect the gravitational field with the ``gravitational charge", for generic compact objects, responsible for the light deflection. To investigate this, rewrite Eqs. \eqref{SCHW} invoking the Tolman-Oppenheimer-Volkov (TOV) prescription,
\begin{equation}\label{metricaTOV}
    ds^2=-\exp^{2\Phi(r)}dt^2+\left(1-\frac{2m(r)}{r}\right)^{-1}dr^2+r^2d\Omega^2,
\end{equation}
in which $\Phi$ and $m(r)$ \emph{exactly represent} the gravitational field and the MS mass \cite{Misner:1973prb}.

Accordingly, switching to Eq. \eqref{metricaTOV}, and computing the refractive index in Eq. \eqref{indicedirifrazione1}, we obtain
\begin{align}
&n_O=\frac{e^{-\Phi(r)}\sqrt{1 - (2 m(r) \sin^2{\sigma})/r}}{\sqrt{1 - 2 m(r)/r}},\label{nOdirettoTOV}
\end{align}
where the MS mass for Eq. \eqref{metricaTOV} clearly reads
\begin{equation}\label{mstov}
m_{MS}=m(r),
\end{equation}
and, so, with arbitrarily $\sigma=0$, from Eq. \eqref{nOdirettoTOV}, we obtain
\begin{equation}\label{mMSvsn0}
   m_{MS}= {1\over2} \left(1 - \frac{e^{-2\Phi}}{n_O^2}\right) r,
\end{equation}
that, combined with Eq. \eqref{mstov}, furnishes,
\begin{equation}\label{intermedia}
    \Phi=-{1\over2}\ln\left[n_O^2\left(1-\frac{2m(r)}{r}\right)\right].
\end{equation}
The MS mass might be non-negative definite. This implies from Eq. \eqref{MSdefinition}, with $\gamma=1$, that
\begin{equation}\label{cclmaf}
    \left|n_O \alpha\right|\geq1\rightleftharpoons     |n_O| \geq e^{-\Phi},
\end{equation}
which agrees with the previous choice $\beta>0$.

Moreover, Eq. \eqref{intermedia} indicates that the potential is ``screened'' by the action of $n_O$, as it appears evident if one is placed at $r\gg r_H$, when the potential will depend on $n_O$ only. Thus, since in general $\frac{2m}{r}\leq1$, for $r\geq r_H$, if $|n_O|\simeq 1$, then $\Phi>0$, whereas if $|n_O|>1$ it is possible that $\Phi$ is positive or negative, depending whether $n_O^2\left(1-\frac{2m(r)}{r}\right)>1$ or not. Hence, the effect of $n_O$ is to \emph{enhance or decrease} the potential magnitude.

Another interesting case is when the gravitational field is small, namely $\Phi\ll1$, in Eq. \eqref{mMSvsn0}, having
\begin{equation}
2\Phi\simeq    1 - n_O^2 \left(1 - \frac{2 m}{r}\right),
\end{equation}
that, again, for large radii, implies that $|n_O|\simeq 1$. Moreover,  in order to have positive MS masses, one needs
\begin{equation}
    n_O^2\geq 1-2\Phi,
\end{equation}
compatible with Eq. \eqref{cclmaf}, since $\Phi\ll1$ implies $    |n_O|\gtrsim \sqrt{1-2\Phi}\simeq 1-\Phi$. However, ensuring $\Phi>0$ implies that if the field is weak, the action of $n_O$ is to \emph{decrease}, alternatively to \emph{un-amplify} the potential, suggesting that amplifying the gravitational effect is a prerogative of intermediate or large fields, not small.

The above relation suggests that \emph{for weak gravitational field} the positivity of the MS mass does not exclude $|n_O|<1$, leading to superluminal effects.

However, more interestingly, it appears that a change of sign of $n_O$ is possible. In fact, demanding $n_O<0$ appears possible from Eq. \eqref{intermedia}, leaving the MS mass positive definite.

We summarize our findings, as follows.

\begin{itemize}
    \item[-] The potential of a given compact object depends on the refractive index that modifies its value.
    \item[-] There is no reason \emph{a priori} to discard the case $n_O<0$ that appears compatible with leaving the MS mass positive and leaves the potential unaltered.
    \item[-] For large radii, the only way to let $\Phi$ vanish consists in assuming $|n_O|\rightarrow1$, \textit{i.e.}, any different asymptotic constraint over $n_O$ would indicate a persistent and non-zero $\Phi$ at infinity.
\end{itemize}

As a general remark, the relationship between the potential, $\Phi$, and $n_O$ indicates that, in the presence of negative refractive indexes and sufficiently strong gravitational fields, it is possible to predict spacetime solutions, and in turn compact objects, whose optical behavior resembles that of \emph{metamaterials} \cite{2013NaPho...7..791M}. Rephrasing it differently, certain spacetimes may act as gravitational sources exhibiting metamaterial-like properties. Since metamaterials are known to be \emph{invisible under specific conditions}, as a consequence of the associated negative refraction \cite{2007NaPho...1..224C}, it is tempting to speculate that the same may occur for spacetimes.

In so doing, a similar mechanism can somehow explain the elusive and non-luminous nature of dark matter, preserving as unique interaction the gravitational one \cite{Arbey:2021gdg}.

To support this conjecture, later in the text we will propose to model dark matter as particle-like configurations arising from spacetimes that exhibit analogies with metamaterials. Specifically, since these solutions mimic the optical properties of metamaterials, we can hereafter refer to them as \emph{gravitational metamaterials}.

\subsection{Sign and properties of refractive index through the redshift definition}

Assuming now to consider the gravitational redshift definition, namely considering $dr=d\Omega=0$, in Eq. \eqref{metricagenerica}, we obtain the proper time,
\begin{equation}\label{tau-ti}
    d\tau=-\alpha dt\,,
\end{equation}
and so, by virtue of the redshift definition,
\begin{equation}\label{zdef}
    z\equiv\frac{\lambda_\infty-\lambda}{\lambda},
\end{equation}
since the wavelength is proportional to the period, \textit{i.e.}, to the proper time, immediately it comes to \cite{Visser:2008cjw,Joshi:2011hb}
\begin{equation}\label{def-z}
    1+z=\frac{1}{|\alpha(r_{source})|}=-g_{00}^{-1/2},
\end{equation}
that holds for observers placed at infinity and for positive $\beta$ and $\alpha$. Interestingly, by comparison of the above expression with Eq. \eqref{indicedirifrazione1}, we obtain
\begin{equation}\label{refractiveredshift}
    n_O=\pm\beta(1+z),
\end{equation}
that vanishes as $z\rightarrow-1$, since according to Eq. \eqref{zdef}, future times imply $\lambda_\infty\rightarrow0$, with $\beta>0$.

This is the standard case, where $n_O$ is positive. Nevertheless, by definition $z+1\geq0$ by construction.

Interestingly, we distinguish two relevant cases, for $\sigma=0$:

\begin{itemize}
    \item[-] Schwarzschild-like coordinates, $\alpha=\beta^{-1}$, with $\gamma=1$. In this case, we end up with
    \begin{equation}\label{soluzionenOscw}
        n_O=\pm(1+z)^2,
    \end{equation}
that implies that $|n_O|\geq1$, with $|n_O|=1$, as $z=0$.

In Schwarzschild coordinates, therefore, the refractive index modulus is always larger than unity.

\item[-] Generic coordinates, hereafter $\alpha\neq\beta^{-1}$, with $\gamma=1$. Here, the possible results are three. The first consists in  $n_O>1$, that is the standard case, explored in classical physics. Here, the signal velocities are \emph{smaller} than the speed of light and the spacetime behaves as a \emph{standard medium}. The second occurrence is $0<n_O<1$. This has no direct explanation in classical physics, albeit $n_O$ is positive definite. Indeed, since $n_O=\frac{c}{v}$, this case is  particularly anomalous, \textit{i.e.}, the signals appear \emph{superluminal} and, only apparently, unphysical. Indeed, the refractive index is associated \emph{with the spacetime itself} and not to a particle within it. Hence, there is no reason to exclude the superluminal case. Finally, the third case, $n_O<0$, \textit{i.e.}, the case claimed above, while dealing with gravitational metamaterials. Its definition is conceptually complicated. The spacetime medium can be reseen as \emph{opaque}, \textit{i.e.}, invisible. This concept is quite different from being ``dark'' and appears recently discussed in solid state physics \cite{2024RvMP...96a5002Y}. It can be visualized requiring that the light is either fully-absorbed by the metric or bent into the opposite direction compared to the standard case.

\end{itemize}

\subsection{Conditions on positive and negative refractive indexes}

A first attempt to obtain a negative refractive index in the weak gravitational field limit has been discussed above, focusing on the TOV spacetime. In this regime, the potential remains unchanged if $n_O<0$, but the optical properties are clearly altered. To better understand how the latter properties change, we can extend the weak-field approximation and ask under what conditions the refractive index can actually become negative.
Thus, starting with $g_{ab}\sim \eta_{ab}
+ h_{ab}$, with $|h_{ab}|\ll1$, and having $g_{ab} \, \frac{\d X^a}{\d t} \, \frac{\d X^b}{\d t},
=0$, we set
\begin{equation} \label{lightray2}
(-1+h_{00}) +
(\delta_{ij} +h_{ij})\, \dot{x}^{i} \, \dot{x}^{j}= 0 \, .
\end{equation}
Afterwards, we split the velocity into a speed and a direction \cite{Boonserm:2004wp},
\begin{subequations}\label{definizionidivelocità}
    \begin{align}
&\dot x^i = || \dot x^i ||
\hat k^i,  \\
&|| \hat k^i||
=\sqrt{ \delta_{ij} \;  \hat k^i\; \hat k^j}=1,
    \end{align}
\end{subequations}
where $\hat k^i$ is a unit 3-vector. Finally, expanding into Taylor series, we obtain
\begin{equation}
 || \dot x^i || =  \sqrt{\frac{1-h_{00}}{1+h_{ij}\, \hat{k}^i \, \hat{k}^j}}
\approx
1-\frac{1}{2} \, h_{00} - \frac{1}{2} \, h_{ij} \, \hat{k}^{i} \,
\hat{k}^{j} + \ldots,
\end{equation}
turning into
\begin{equation}
n(\hat k) = \frac{1}{ || \dot x^i ||} \approx
1 + \frac{1}{2} \, h_{00} +  \frac{1}{2} \, h_{ij} \,
\hat{k}^{i} \, \hat{k}^{j} + \, \mathcal{O}(h_{ab}^2),
\end{equation}
that, compared with Eq. \eqref{def-z}, furnishes
\begin{equation}
z\simeq \frac{h_{00}}{2}.
\end{equation}

An interesting remark is that, in the above refractive
index prescription, the light beam travels in the direction $\hat k$. Hence, to obtain a negative refractive index, in the absence of an imaginary refractive index, one is forced to suppose that \emph{the light beam travels in the opposite direction given by $\hat k$}. To figure this out, let us rewrite the first relation in Eq. \eqref{definizionidivelocità} as
\begin{equation}\label{dotxi}
    \dot x^i = -|| \dot x^i ||
\hat k^i,
\end{equation}
to obtain the opposite direction, enabling to have a negative refractive index. Following the recipe of metamaterials in solid state physics, the result in Eq. \eqref{dotxi} may suggest that a negative refractive index means that the wave vector and the Poynting vector, indicating the direction in which energy flows, point in opposite directions.

Accordingly, \emph{negative refraction suggests that light bends ``the other way", when passing through the metric inducing $n_O<0$}.

For the sake of clearness, the result in Eq. \eqref{dotxi} appears quite anomalous and, physically, can be attributed to the fact that we are dealing with a \emph{single spacetime only} that exhibits a negative refractive index. Suppose, instead, that we can have multiple spacetimes, connected to each other, that behave as \emph{layers}. In this case, an effective negative refractive index can be constructed, similar to the construction made in the laboratory of invisible structures \cite{2010ApPhL..96l1910J}. In particular, matching spacetimes can be accounted through the Israel-Darmois matching conditions \cite{Israel:1966rt}, which, however, are beyond the purpose of this work.

In addition, the physical interpretation of the refractive index, at this stage, appears to be only partially explained, since the electromagnetic field has not been included yet in both the direct and indirect treatments. In the next section, its inclusion will shed light on its connections with spacetime media.

\section{Gravitational refractive index incorporating the electromagnetic field}\label{sezione3}

In this section, we analyze two main points,
\begin{itemize}
    \item[-] the physical meaning of $n^{inh}$ and $n^{anis}$,
    \item[-] the meaning of every term of the metric inside $n_O$.
\end{itemize}

It is a consolidate fact that the refractive index was originally derived from electromagnetic properties of media. In general, in fact, all gravitational compact objects may reflect or absorb electromagnetic radiation, turning into the great advantage to infer information about their structure, composition, temperature, age, dynamics, distance and so on.

However, within the above treatment, $n_O$ did not incorporate the electromagnetic field and, accordingly, the underlying definitions, direct and indirect, were not fully exhaustive.

We show here that, once the electromagnetic field is included, the purely geometrical refractive index predicts an anisotropic part that can be matched with our findings in Eq. \eqref{anis}. Phrasing it differently, the direct and indirect strategies predict $n^{inh}$, which turns out to be the same once the electromagnetic field is involved, albeit under precise circumstances.

First, we notice that for a spherically-symmetric spacetime, the refractive index is clearly $r$-dependent, namely inhomogenous, and this appears as a \emph{general fact} that \emph{does not depend on the refractive index definition}, \textit{i.e.}, neither including nor excluding the electromagnetic fields.

Physically speaking, the refractive index depends upon the choice of the radial coordinates and, consequently, choosing the most appropriate coordinate transformation may help in describing the spacetime as a medium in its possible simplest form. This is particularly evident if one invokes the analogy between \emph{anisotropic crystals} and spacetimes \cite{Leonhardt:2006ai}.

Thus, consider the  electromagnetic field in curved spacetime
\begin{equation}\label{DBEH}
D^i=\varepsilon^{ij}E_j-(\Gamma \times \mathbf{H})^i, \quad
B^i=\mu^{ij}H_j+(\Gamma \times \mathbf{E})^i,
\end{equation}
where $\mathbf{D}$ and $\mathbf{E}$ whereas  $\mathbf{B}$, and
$\mathbf{H}$ are the electric and magnetic fields in vacuum and in the media, respectively, with an asymmetric vector,  $\mathbf{\Gamma}$, given by $\Gamma_i = -  \frac{g_{0i}}{g_{00}}$.

The fields are related to each other through the permittivity, $\varepsilon^{ij}$, and permeability tensors, $\mu^{ij}$, namely \cite{Fernandez-Nunez:2015qva}
\begin{equation}
\varepsilon^{ij}=\mu^{ij}=\sqrt{-g}\frac{g^{ij}}{g_{00}}.
\label{eq:perm}
\end{equation}
Quite remarkably,
$\varepsilon^{ij}$ and $\mu^{ij}$ turn out to be symmetric and $\Gamma=0$ when $g_{0i}=0$. This case includes the \emph{time-reversal symmetry}, that appears analogous to what happens for materials. This case is addressed for diagonal metrics. In fact, the discrete symmetry
$t\rightarrow-t$ is fulfilled through the condition
$g_{0i} =0$, which holds in our case, as evident from Eq.~\eqref{metricagenerica}. Moreover, we remark that it is particularly important to avoid off-diagonal terms in the refractive index, for at least two reasons: 1) it results in a simpler mathematical structure for the given solution, 2) it facilitates working with a more straightforward representation of the medium.

Unfortunately, this hypothesis is stringent and does not occur in all the media. Hence, to model an electromagnetic medium using a spacetime, one can invoke the same coordinate transformations, presented in Eq. \eqref{eq:r-rho} \cite{review,Leonhardt:2006ai}, bu including here, for generality $g_{ij}\neq0$, but preserving the time reversal symmetry, say $g_{0i}=0$, in Eq. \eqref{metricagenerica}.

Thus, using the coordinate set, $(t,r,\theta,\phi)$, we write in Eq. \eqref{metricagenerica},  $\alpha^2=g_{00}$, $\beta^2=g_{11}$ and $\gamma^2=g_{11}\eta^2$, with $\eta=\eta(r)$ an auxiliary function, recasting Eqs. \eqref{lam}-\eqref{f2} to give $n_O$ as
\begin{equation}\label{nOgenerico1}
|n_O|\delta^{ij}=\frac{\beta}{\alpha}\Big[\delta^{ij}-\left(1-\eta^2\right)\frac{x^{i}x^j}{r^2}\Big],
\end{equation}
that, again, following the procedure pursued in Eqs. \eqref{inh}-\eqref{anis}, can be split into \cite{Sen:2010zzf}
\begin{subequations}\label{inheanisbis}
    \begin{align}
n_O&\equiv\sqrt{\epsilon^{ij}\mu^{ij}}=n_O^{inh}+n_O^{anis},\\
n_O^{inh}&=\frac{\beta}{\alpha},\\
n_O^{anis}&=-\frac{\beta}{\alpha}\left(1-\frac{\gamma^2}{\beta^2}\right)\frac{x^{i}x^j}{r^2},
    \end{align}
\end{subequations}
where $r=\sqrt{x^2+y^2+z^2}$, $n_O^{inh}$ and $n_O^{anis}$ are respectively the inhomogeneous and anisotropic parts of the refractive index, $\delta^{ij}$ is the Kronecker delta and $x^{i}x^j$ are written in terms of Cartesian coordinates, $(x,y,z)$.

The inhomogeneous expression coincides with Eq. \eqref{inh}, when $i=j$, whereas the anisotropic term is \emph{in form} compatible with Eq. \eqref{anis}, coinciding \emph{de facto} with it when the conditions in Eqs. \eqref{smallconditions} are fulfilled and, again, if $i=j$. Hence, the introduction of Eqs. \eqref{DBEH} implies a different anisotropic refractive index that includes off-diagonal terms, if $g_{ij}\neq0$ and generalizes the $n_O$ definition.

Hence, the advantage of introducing the electromagnetic field lies on the fact that anisotropy emerges naturally from the term $\sim x^i x^j$ in Eq. \eqref{inheanisbis}. This term includes off-diagonal contributions, corresponding to anisotropic components in crystals, \textit{i.e.}, following \emph{de facto} the analogy between media and spacetimes. However, such anisotropic effects can even vanish in the limit $\eta \to 1$, even when $i \neq j$.

In addition, the refractive index, $n_O$, is in general a tensor and,  so, the auxiliary function $\eta$ bids the anisotropic part to behave, increasing or decreasing, and this justifies the adjective ``anisotropic", used in Eq. \eqref{anis} while describing the two refractive index terms, arising when one adopts general relativity only.

Nevertheless, when
\begin{equation}\label{conditio}
    \gamma=\beta\rightarrow \eta=1,
\end{equation}
our \emph{conformally flat metric}, Eq. \eqref{metricagenerica2conforme}, furnishes  $n_O\equiv n_O^{inh}$, providing
\begin{equation}
\varepsilon^{ij} = \mu^{ij}\rightarrow |n_O|= \frac{\beta}{\alpha},
\label{nOgenerico1}
\end{equation}
giving a \emph{inhomogeneous but isotropic spacetime medium}.

The condition \eqref{conditio} is therefore the way to isotropize the refractive index, \textit{i.e.}, to isotropize the medium associated with the spacetime and, so, the function $\eta$ represents an \emph{anisotropic factor}, since it quantifies the deviation from isotropy. Nevertheless, if $\eta \neq 1$ the refractive index contains off-diagonal terms, if $i\neq j$.

From the definitions of the refractive indices, both with and without electromagnetic fields, it is evident that they coincide only in the inhomogeneous component, whereas the anisotropic refractive index changes its form when the electromagnetic field is included.

However, it appears evident that its expression is analogous to that obtained in general relativity for small $\sigma$. Hence, spacetimes with $g_{ij}\neq0$ would furnish the same finding, mimicking the refractive index tensor, obtained using the electromagnetic fields as a geometric effect, for very small refractive angles, say $\sigma\ll1$.

To this end, future measurements can be imagined to seek both inhomogeneous and anisotropic contributions that differently emerge, according to the kind of spacetime that models a given compact object. This fact may, in turn, provide insights into the structure of the examined object that generates the gravitational interaction itself.

\section{Optical properties of spacetimes}\label{sezione4}

The existence of a gravitational optical index, that mimes the index found in Eq. \eqref{nOgenerico1},  implies that, for \emph{any spacetime} the corresponding  optical properties strongly depend on the form of the metric itself and \emph{can mime the structure of different kinds of media}.

In addition, in view of the great interest in regular solutions, \emph{we wonder whether it is possible to distinguish regular from singular spacetimes} using $n_O$. In the same way, solutions that do not exhibit horizons, \textit{i.e.}, naked singularities and/or wormholes, \emph{can in principle be distinguished from their optical properties from regular and singular spacetimes}. In this respect, in the following we work out the three distinct cases, \textit{i.e.}, standard black holes, regular black holes,  and solutions with no horizons, among which naked singularities.

\subsection{Refractive indexes of black holes}

In Eq. \eqref{SCHW}, the simplest Schwarzschild solution has been considered. Other quite relevant cases are, in addition, the Schwarzschild-de Sitter metric, where we add a constant vacuum energy contribution (de Sitter phase), $\Lambda$, and the Reisser-Nordstr$\ddot{\rm o}$m metric, introducing an electric charge, $Q$. We thus write the Schwarzschild-de Sitter metric as,
\begin{subequations}\label{SdeSitter}
    \begin{align}
&\alpha=\sqrt{1-\frac{2M}{r}-\frac{\Lambda r^2}{3}}=\beta^{-1},\\
&\gamma=1,
    \end{align}
\end{subequations}
and the Reisser-Nordstr${\rm \ddot o}$m metric as
\begin{subequations}\label{RNmetrica}
    \begin{align}
&\alpha=\sqrt{1-\frac{2M}{r}-\frac{Q^2}{r^2}}=\beta^{-1},\\
&\gamma=1.
    \end{align}
\end{subequations}
The Schwarzschild-de Sitter horizons are
\begin{equation}\label{rhSdS}
    r_{H}^{\pm} \equiv
    \frac{2}{\sqrt{\Lambda}}\cos\left[\frac{1}{3}\cos^{-1}\left(3\sqrt{\Lambda}M\right)\pm\frac{\pi}{3}\right],
\end{equation}
where $r_H^+<r_H^-$, while the third solution is negative, \textit{i.e.}, $r=r_0\equiv-(r_H^++r_H^-)$, and clearly unphysical since one cannot extend the coordinate range beyond the curvature singularity placed at $r=0$. For the  Schwarzschild-de Sitter metric, we limit ourselves to $3\sqrt{3}M<\frac{3}{\Lambda}$.

For the Reisser-Nordstr$\ddot {\rm o}$m metric, we have again two horizons,
\begin{align}\label{horinRN}
r_H^{RN}=M\pm\sqrt{M^2-Q^2}.
\end{align}

From Eq. \eqref{indicedirifrazione1}, the refractive indexes become
\begin{subequations}
    \begin{align}
        &n^{SdS}_{O}=\frac{r \sqrt{9 - [3 (6 M + r^3 \Lambda) \sin^2\sigma]/r}}{3 r-6 M - r^3 \Lambda},\\
        &n^{RN}_{O}=\frac{r^2 \sqrt{1 + [(Q^2 - 2 M r) \sin^2\sigma]/r^2}}{Q^2 + r (r-2 M)}.
    \end{align}
\end{subequations}
The two solutions have more than one free parameter, as for the Schwarzschild metric. Consequently, a direct comparison among the refractive indexes can be computed by evaluating the  departures from the Schwarzschild case and to clarify the role of $\Lambda$ and $Q$ in modifying $n_O$.

To do so, having $n_O^S$ the refractive index in Eq. \eqref{nS} and assume conventionally $\sigma\ll1$, we can introduce the following definition,
\begin{eqnarray}\label{percentualeBH}
    \delta n_O\equiv\frac{n_O^i-n_O^S}{n_O^i},
\end{eqnarray}
with $i=\{RN,SdS\}$, referring to the two above black holes. By virtue of Eq. \eqref{percentualeBH}, and considering small $Q$ and $\Lambda$, we obtain
\begin{subequations}
    \begin{align}
        &\delta n^{SdS}_{O}=\frac{r^3\Lambda}{3(r-2M)}+\mathcal O(\sigma^2),\\
             &\delta n^{RN}_{O}=-\frac{Q^2}{r(r-2 M)}+\mathcal O(\sigma^2).
    \end{align}
\end{subequations}

\begin{figure}
    \centering
    \includegraphics[width=8.3cm,height=5.2cm
]{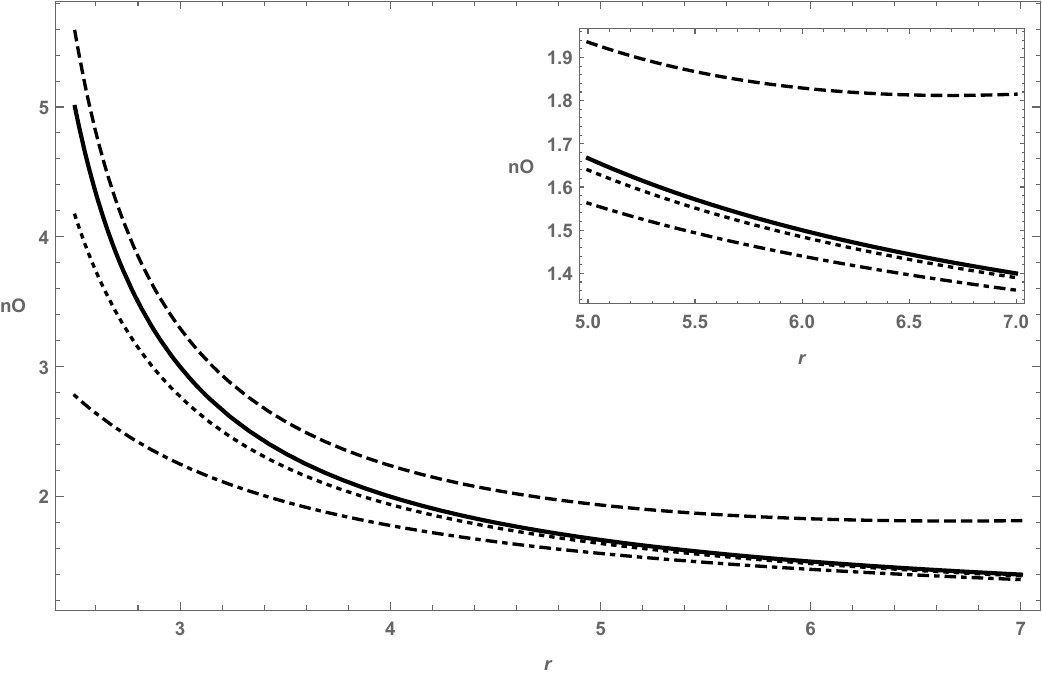}
    \caption{Plots of $n_O$ for the black hole solutions. Precisely, the thick black line corresponds to the Schwarzschild black hole, while the dashed and dotted lines are respectively  the Schwarzschild-de Sitter and Reisser-Nordstr$\ddot {\rm o}$m spacetimes, with indicative unitary mass and  $Q=0.5$, $\Lambda=0.01$ and $\sigma=0$. Remarkably, close to the horizon, $r_H\sim 2M$, the refractive index tends to infinity as expected, while at larger distances it decreases to zero. Quite interestingly, the extreme case $M=Q$ is displayed in the dot-dashed curve, corresponding to the maximal departure of the Reissner-Nordstr$\ddot {\rm o}$m metric from the Schwazschild case. }
    \label{deltaindiciperBHs}
\end{figure}

Immediately, for $r>2M$, it is evident that $\delta n_O^{RN}<0$ and $\delta n_O^{SdS}>0$ for $\Lambda>0$. Hence, we end up with the behaviors of $\delta n_O$ for the black hole solutions, lying within an absolute difference percentage being roughly $\gtrsim 20\%$ for both Reissner-Nordstr$\ddot {\rm o}$m and Schwarzshild-de Sitter cases, except close to the horizons, where it increases significantly. Nevertheless, the shapes of  $n_O$ are also reported in Fig. \ref{deltaindiciperBHs}, considering the interesting extreme case,  $M=Q$, in the Reissner-Nordstr$\ddot {\rm o}$m solution and highlighting the significant differences with the case $M\neq Q$. The plots have been obtained considering the constraint imposed by the Schwarzschild-de Sitter horizons, in Eq. \eqref{rhSdS}.

Accordingly, the action of the charge is to \emph{decrease} the refractive index, since physically it acts to decrease the effective MS mass. Moreover, The extremal case implies the \emph{maximum decrease possible} in proximity of the horizon. The action of the de Sitter phase is, instead, to \emph{increase} the refractive index, namely a positive de Sitter phase acts by increasing the mass of the MS.

In the overall treatment, the signs of $\alpha$ and $\beta$, in Eqs. \eqref{SdeSitter}-\eqref{RNmetrica}, have been conventionally taken positive. Negative refraction can be obtained, then, ensuring $\alpha<0$, as explained above.

\subsection{Refractive indexes of regular black holes}

Analogously to black holes, the regular black hole solutions can be compared with the standard Schwarzschild scenario. We here select four popular regular black holes \cite{Lan:2023cvz}, \textit{i.e.}, the Bardeen spacetime,
\begin{subequations}\label{Bardeen_func}
\begin{align}
&\alpha=\sqrt{1-\frac{2Mr^2}{(r^2+q^2)^{3/2}}}=\beta^{-1},\\
&\gamma=1,
    \end{align}
\end{subequations}
defined by a topological charge, $q$; the Hayward metric,
\begin{subequations}\label{Hayward_func}
    \begin{align}
&\alpha=\sqrt{1-\frac{2Mr^2}{r^3+2a^2}}=\beta^{-1},\\
&\gamma=1,
    \end{align}
\end{subequations}
with $a^2=M\Lambda^{-1}$, having  $\Lambda$ under the form of a de Sitter phase, attributed to vacuum energy contribution; the Dymnikova and Fang-Wang spacetimes, respectively
\begin{subequations}\label{Dymnikova_func}
    \begin{align}
    &\alpha=\sqrt{1-\frac{4M}{\pi r}\left[\arctan\left(\frac{r}{l_{\rm D}}\right)-\frac{r \ l_{\rm D}}{(r^2+l_{\rm D}^2)}\right]}=\beta^{-1}\,,\\
    &\gamma=1,
\end{align}
\end{subequations}
and
\begin{subequations}\label{FW_func}
    \begin{align}
    &\alpha=\sqrt{1-\frac{2Mr^2}{(r+l_{\rm FW })^{3}}}=\beta^{-1}\,,\\
    &\gamma=1,
\end{align}
\end{subequations}
defined by the characteristic lengths, $l_D$ and $l_{FW}$.

These scenarios  asymptotically resemble the Schwarzschild solution, but close to the cores, their behavior is regular through the addiction of either topological charge, or cosmological constant and length contributions \cite{Luongo:2023aib,Carballo-Rubio:2025fnc,Buoninfante:2024oxl}.

Accordingly, we immediately obtain
\begin{subequations}\label{nORBH}
    \begin{align}
        n_O^{B}&=\frac{\sqrt{1 - \frac{2 M r^2 \sin^2\sigma}{(q^2 + r^2)^{3/2}}}}{1 - \frac{2 M r^2}{(q^2 + r^2)^{3/2}}},\\
        n_O^{H}&=\frac{\sqrt{1 - \frac{2 M\Lambda r^2 \sin^2\sigma}{2 M + r^3\Lambda}}}{1 - \frac{2 M\Lambda r^2}{2 M + r^3\Lambda}},\\
        n_O^{D}&=\frac{\sqrt{1 + (
 4 M/\pi) \left(\frac{l_D}{l_D^2 + r^2} - \frac{\tan^{-1}(r/l_D)}{r}\right) \sin^2\sigma}}{1 + (
 4 M/\pi) \left(\frac{l_D}{l_D^2 + r^2} - \frac{\tan^{-1}(r/l_D)}{r}\right)},\\
        n_O^{FW}&=\frac{\sqrt{1 -\frac{2 m r^2 \sin^2\sigma}{(l_{FW} + r)^3}        }}{1 - \frac{
  2 M r^2}{(l_{FW} + r)^3}}.
    \end{align}
\end{subequations}

In so doing, close to the de Sitter core for each solution, from Eqs. \eqref{nORBH}, we find
\begin{subequations}\label{RBHcores}
    \begin{align}
       n_o^{B}(r\simeq0)&\sim1+\frac{M}{q^3}(1+\cos^2\sigma)r^2,\\
       n_o^{H}(r\simeq0)&\sim 1+\frac{M}{2a^2} (1+\cos^2\sigma)r^2,\\
       n_o^{D}(r\simeq0)&\sim1+\frac{4M}{3\pi l_D^3}(1+\cos^2\sigma)r^2,\\
       n_o^{FW}(r\simeq0)&\sim1+\frac{M}{l_{FW}^3}(1+\cos^2\sigma)r^2.
    \end{align}
\end{subequations}

\begin{figure}
    \centering
    \includegraphics[scale=0.5
]{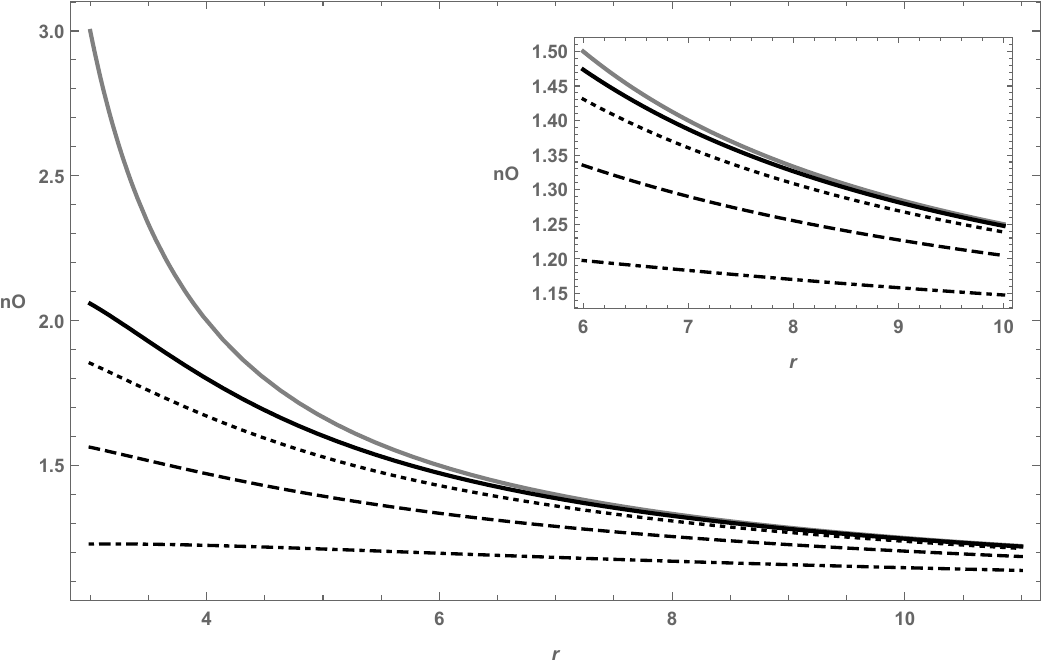}
    \includegraphics[scale=0.5
]{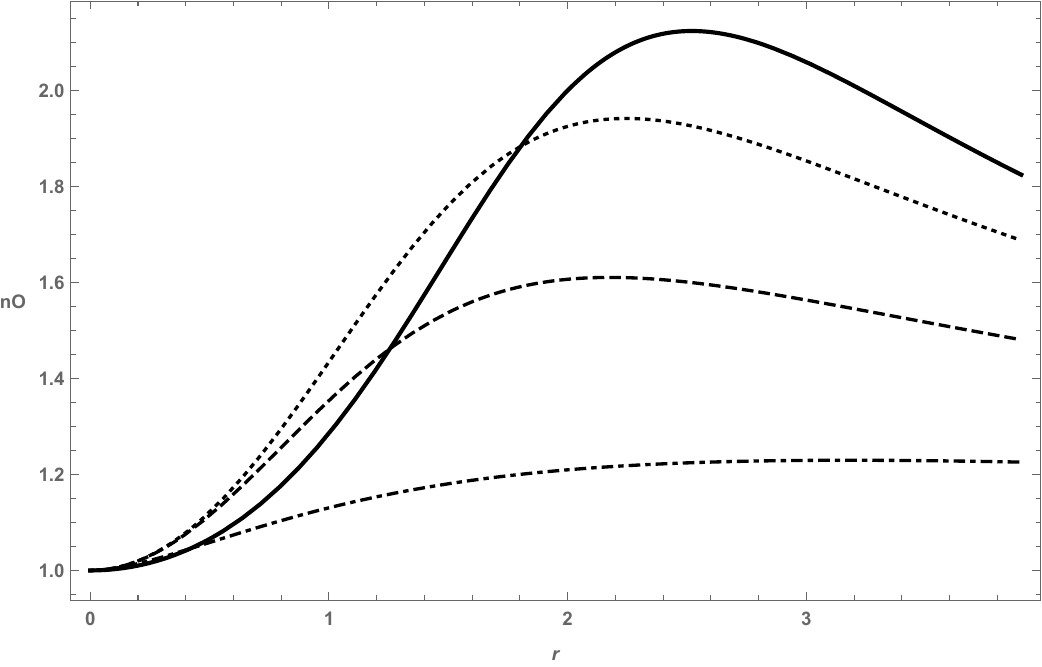}
    \caption{Plots of the refractive indexes for the regular black hole solutions, as reported in Eqs. \eqref{nORBH}; (top figure). Plots of $n_O$, compared with the  Schwarzschild black hole (thick gray line), beyond the horizon. The black thick, dotted, dashed and dot-dashed lines are respectively the Hayward, Bardeen, Dymnikova and Fang-Wang regular black holes; (bottom figure). Plot of $n_O$ within the horizon for all the solutions, excluding Schwarzschild that would tend to negative values. In all the plots,  we adopted an indicative unitary mass and arbitrarily  $\Lambda=\frac12$ and  $\sigma=0$.
    At $\sigma=0$ and $a^2\equiv M\Lambda^{-1}$, comparing the Hayward de Sitter core with the other three behaviors in Eqs. \eqref{RBHcores}, the values of $q,l_D$ and $l_{FW}$ can be expressed in terms of the Hayward de Sitter phase, $\Lambda$, namely $q=\left(\frac{2}{\Lambda}\right)^{\frac13}$, $l_{FW}=\left(\frac{2}{\Lambda}\right)^{{1\over3}}$ and $l_D=\left(\frac{8}{3\pi\Lambda}\right)^{1/3}$.
    The main departures from the Schwarzschild solution occur close to the horizon, as expected, while asymptotically the solutions tend to match.}
    \label{indiciperRBHs}
\end{figure}

The qualitative behaviors of such optical indexes are reported in Fig. \ref{indiciperRBHs}, where we specialize each figure for $n_O>0$. Quite remarkably, before the horizons, the refractive indexes may be positive and, in any cases, with the same signs than those evaluated beyond the horizons. This appears very different with respect to the black hole case, where $n_O$ is supposed to change sign once crossing the horizon \cite{optical}.

Conversely, the departures from the regular metrics and the Schwarzschild spacetime are
\begin{subequations}
    \begin{align}
        &\delta n^{B}_{O}=-\frac{3 M q^2}{r^2 (r-2 M)}+\mathcal O(\sigma^2),\\
        &\delta n^{H}_{O}=-\frac{4 a^2 M}{(r-2M) r^3}+\mathcal O(\sigma^2),\\
        &\delta n^{D}_{O}=-\frac{8 l_D M}{\pi r(r-2 M)}+\mathcal O(\sigma^2),\\
        &\delta n^{FW}_{O}=-\frac{6 l_{FW} M}{r(r-2 M)}+\mathcal O(\sigma^2),
    \end{align}
\end{subequations}
computed for simplicity at $\sigma\ll1$, using i=\{B;H;D;FW\} in Eq. \eqref{percentualeBH}. The departures are of second order in $q$ and $M\Lambda^{-2}$, while first order in $l_{D}$ and $l_{FW}$ and \emph{appear negative for all the metrics}. Their behaviors are reported in Fig. \ref{indiciperRBHs}. Generally, the departures differ particularly, indicating that each regular solution is extremely different than others. Quite remarkably, the closest to well-known singular solutions are the Bardeen and Hayward solutions, whose departures appear comparable with those computed in the case of Reisser-Nordstr$\ddot {\rm o}$m and Schwarzschild-de Sitter black holes.

However, again, for regular metrics, it appears possible to find negative refractive indexes, even beyond the horizon, ensuring $\alpha<0$, as well as black holes.

\subsection{Refractive indexes of naked singularities and regular solutions with no horizon}

The cosmic censorship conjecture forbids the existence of naked singularities \cite{Malafarina:2022wmx}. However, their presence could, in principle, be related to repulsive gravity effects \cite{Luongo:2023xaw} and other intriguing phenomena, such as the progenitors of gamma-ray bursts \cite{Antia:1998cx,Eichler:1989ve}, among others.

In general, above we discussed about at least two notable optical properties. The first is evident in the strong increase of $n_O$ close to horizons. The second refers to the interior structure: crossing the horizon results in a sign reversal of the lapse and shift functions. For regular black holes this does not happen, as for black holes, instead.

For naked singularities, however, there is no horizon cross and, accordingly, a naked singularity exhibits smoothly evolving, either positive or negative, refractive indexes.

For example, the prototype of a naked singularity arises from the mass reversal, $M\rightarrow-M$. While this framework reveals interesting regions of repulsive gravity \cite{Luongo:2010we}, it suffers from severe instability issues, see, e.g., Ref. \cite{Gleiser:2006yz}. Nonetheless, it serves as a benchmark for comparing other naked singularity models.

Another remarkable naked singularity can be obtained, considering $M<Q$ in Eq. \eqref{horinRN}, providing a Reisser-Nordstr${\rm \ddot o}$m spacetime with no horizons.

In this respect and more remarkably departing from the above recipe, a very simple solution without horizons may be obtained once considering the anti-de Sitter (AdS) solution or the classes of AdS solutions.

These have recently acquired great importance in view of the correspondence with conformal field theories, named AdS/CFT correspondence \cite{Maldacena:1997re,Hubeny:2014bla}.

\begin{figure}
    \centering
    \includegraphics[scale=0.5
]{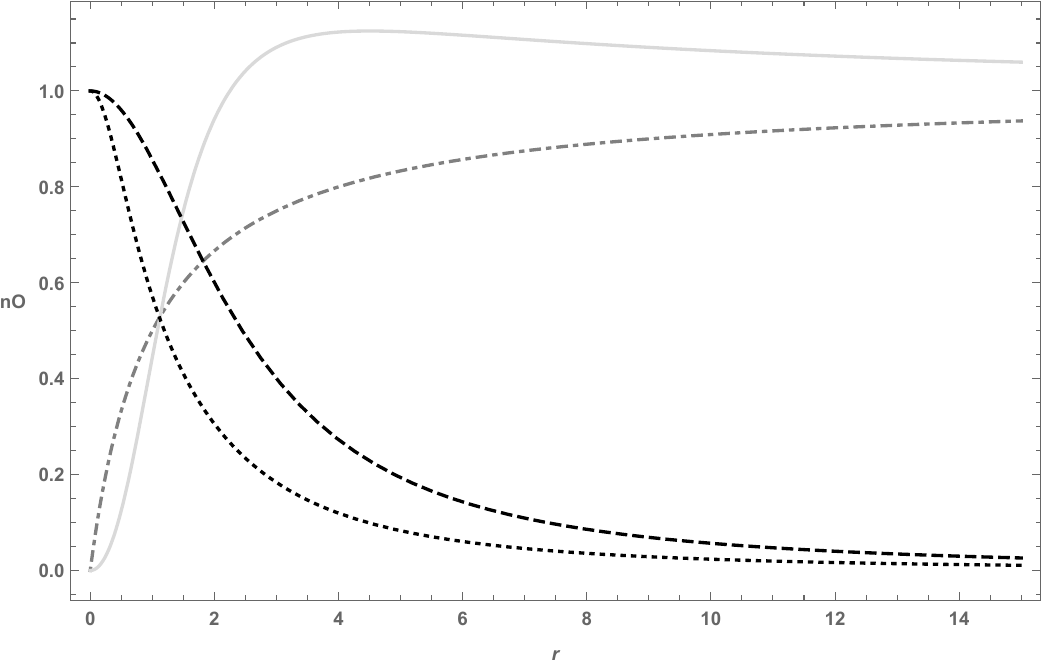}
    \caption{Plots of the refractive indexes for solutions with no horizon. The AdS and quasi AdS solutions are displayed in dashed and dotted black lines, respectively. Indicatively, we also reported the Schwarzschild black hole with $M\rightarrow-M$, in gray as dot-dashed curve, with $M=-0.5$. The action of $Q$ is portrayed in the light gray curve, where we plot the Reisser-Nordstr{\rm $\ddot{o}$}m solution with $Q=3M$ and $M=0.5$. The behaviors of each solution is particularly different than singular and regular solutions, indicating superluminal regions. The indicative values, here used for the plots, are  $\Lambda=0.5$ and $k_0=4\Lambda$. The AdS and quasi AdS solutions appear quite similar, as expected,  conversely to the Schwarzschild solution with negative mass whose concavity appears quite different, as a consequence of $M$. }
    \label{antideSitter}
\end{figure}

First, we may focus on the AdS solution,
\begin{subequations}\label{AdS}
    \begin{align}
&\alpha=\sqrt{1+\frac{\Lambda r^2}{3}}=\beta^{-1},\\
&\gamma=1,
    \end{align}
\end{subequations}
with $\Lambda>0$, providing no horizons. The refractive index, in Eq. \eqref{indicedirifrazione1}, for the AdS solution gives
\begin{equation}\label{indicedirifrazioneAdS}
n_O^{AdS}=\frac{\sqrt{9 + 3 r^2 \Lambda \sin^2{\sigma}}}{3 + r^2 \Lambda},
\end{equation}
whereas a possible generalization, as alternative candidate to AdS scenarios, has been proposed in Ref. \cite{miametrica}, constructed by
\begin{subequations}
    \begin{align}
&\alpha=\sqrt{1+\frac{\Lambda r^2}{3}},\\
&\beta=\sqrt{\left(1+\tfrac\Lambda 3  r^2\right)\frac{1+k_0 r^2 }{1+2 \Lambda  r^2/3}},\\
&\gamma=1,
    \end{align}
\end{subequations}

\noindent leading to

\begin{equation}\label{indicedirifrazioneqAdS}
n_O^{{\small quasi-AdS}}=\mathcal N\sqrt{1-\frac{3r^2(\Lambda+k_0(1+r^2\Lambda/3))\sin^2\sigma}{(1+k_0r^2)(1+r^2\Lambda/3)}},
\end{equation}
with $\mathcal N\equiv \sqrt{(1-2r^2\Lambda/3)(1+k_0r^2)^{-1}}$,
where $k_0$ is a free constant, whose limiting case, $k_0=\frac{2}{3}\Lambda$, yields Eq. \eqref{AdS}. Analogously, we here consider positive values of $\Lambda$. The behaviors of Eqs. \eqref{indicedirifrazioneAdS} and \eqref{indicedirifrazioneqAdS}
 are reported in Fig. \ref{antideSitter}, together with the Schwarzschild solution exhibiting negative mass and Reisser-Nordstr${\rm \ddot o}$m with $M<Q$.

Differently of regular black holes, the behaviors at $r\simeq0$ increase, indicating the presence of the singularity.

The $\delta n_O$ for the two naked singularities are effectively much different than all the solutions, involved so far. The main reason consists in the absence of mass, as in all the other solutions.

A summary related to the properties of the kinds of solutions, above cited, namely including black holes, regular metrics and naked singularities, is reported in Table \ref{tab:sommario}.

\begin{table*}
\small
\begin{tabularx}{\textwidth}{p{4cm} p{5cm} p{5cm}}
\centering Kind of solution & \centering Material behaved spacetime & \hspace{0.4cm}Metamaterial behaved spacetime  \\
\hline
\begin{center} Black hole  \end{center} &
\begin{itemize}
\item[$\blacktriangleright$] The refractive index diverges (tends to one) at $r\rightarrow r_H$ (at $r\gg r_H$) in Schwarzschild coordinates.
\item[$\blacktriangleright$] The refractive index \emph{changes} sign as $r< r_H$. It diverges from left in the singularity, $n_O(0)\rightarrow-\infty$.
\end{itemize} &
\begin{itemize}
\item[$\blacktriangleright$] Same as material but with opposite signs. The asymptotic behavior at $r\gg r_H$ defines an upper limit for $n_O$.
\item[$\blacktriangleright$] The refractive index is negative for $r>r_H$ and positive $r< r_H$, changing sign as for material behaved spacetimes, but diverging from right in the singularity.
\end{itemize}\\
\hline
\begin{center}Regular black hole  \end{center} &
\begin{itemize}
\item[$\blacktriangleright$] Quite differently from black holes, the refractive index can here reach a maximum value (in principle, measurable) as $r\rightarrow r_H$ and decreases when $r<r_H$, preserving the sign.
\item[$\blacktriangleright$] The refractive index \emph{does not change} sign as $r< r_H$. It converges to unity in both the singularity, $n_O(0)\rightarrow1$ and large radii, $r\gg r_H$.
\end{itemize} &
\begin{itemize}
\item[$\blacktriangleright$] In analogy to the material behaved solutions, the refractive index can here reach a minimum value as $r\rightarrow r_H$ (possibly indirectly detectable) and then increases when $r<r_H$.
\item[$\blacktriangleright$] The refractive index \emph{does not change} sign as $r< r_H$. At $r=0$ and $r\gg r_H$, $n_O(0)\rightarrow-1$, becoming perfectly opaque.
\end{itemize}\\
\hline
\begin{center} Solutions with no horizon \end{center}&
\begin{itemize}
\item[$\blacktriangleright$] Regardless the absence of mass, e.g., for the AdS and quasi-AdS solutions, the refractive index tends to zero into $r=0$, while being in some regions, or everywhere, smaller than unity, indicating a \emph{superluminal effect}.
\item[$\blacktriangleright$]  The refractive index does not necessarily tends to unity at infinity, indicating a possible unphysical behavior.
\end{itemize} &
\begin{itemize}
\item[$\blacktriangleright$] The refractive index tends to its maximum as $r\rightarrow0$, while being in some regions, or everywhere, larger than $-1$, indicating some sort of \emph{optical repulsive effect}.
\item[$\blacktriangleright$] Analogous unphysical behavior at infinity if compared with material behaved spacetimes.
\end{itemize}\\
\hline
\end{tabularx}
\caption{Comparison among the different spacetimes, split into classes of black holes, regular solutions and naked singularities. For each of them, it is reported in detail the possibility to have both material and metamaterial behaviors. In all the classes, we ensure $\gamma$ to be a smooth and regular function, \textit{i.e.}, neglecting any sort of singularities or discontinuities on it. Moreover, the term $\frac{\gamma^2}{\beta^2}$ appearing in the anisotropic $n_O$, \textit{i.e.}, Eq. \eqref{anis}, is also assumed to be smaller than unity. The results do not depend on selecting a particular value of $\sigma$.}
\label{tab:sommario}
\end{table*}

\section{Snell’s law and radial gravity approximation}\label{sezione5}

In this section, we investigate the consequences of our treatments using Snell's law. We therefore fix constraints on black holes and regular solutions, with or without horizons, trying to distinguish their effects.

The refractive indexes in two spacetimes are intimately related to the direction of light rays in different gravitational configurations.

So, assuming that no discontinuity occurs between two spacetime media, we require
\begin{eqnarray}\label{snellbasico}
n_{O,1}\sin(\sigma_1) = n_{O,2}\sin(\sigma_2+\delta \phi),
\end{eqnarray}
where, $n_{O,1} = n_{O,1}(r_1, \sigma_1)$ and $n_{O,2} = n_{O,2}(r_2, \sigma_2)$ represent the refractive indexes associated with the light rays (incident and refracted) against the outward normal, defined through the deflection angles, $\sigma_1$ and $\sigma_2$.

The discrepancy between $\sigma_1$ and $\sigma_2$ is $\delta\phi$, representing the incremental angle, generated by $r$, when it rotates to acquire the values $r_1$ and $r_2$.

An astute procedure can be applied to transform Eq. \eqref{snellbasico} into a differential equation by
\begin{eqnarray}
&&n_{O,1}\frac{\sin(\sigma_2 + \delta\phi) - \sin(\sigma_1)}{\sigma_2 + \delta\phi - \sigma_1}\,(\sigma_2 + \delta\phi - \sigma_1)\nonumber\\
&&= -\frac{n_{O,2} - n_{O,1}}{\delta\phi}\sin(\sigma_2 + \delta\phi)\,\delta\phi,
\end{eqnarray}
by ensuring $\sigma_2 -\sigma_1\rightarrow 0$ and simultaneously  $\delta\phi\rightarrow0$. So, it is straightforward to obtain,
\begin{eqnarray}\label{espressionegen}
\sigma^\prime\equiv\frac{d\sigma}{d\phi}= -1 - \frac{r}{n_O}\frac{dn_O}{dr}\frac{\tan\sigma\,dr}{rd\phi}.
\end{eqnarray}

At this stage, it is not clear \emph{a priori} how to compute $\frac{dn_O}{d\phi}$, since we do not explicitly have $n_O=n_O(\phi)$. Nonetheless, for very small angles, one can approximate the variation of $\phi$ by \cite{optical}
\begin{equation}\label{interm}
    \frac{d\phi}{dr}=\frac{\tan\sigma}{r},
\end{equation}
and, so, combining Eq. \eqref{espressionegen} with Eq. \eqref{interm}, we obtain
\begin{eqnarray}\label{eq:9}
\frac{d\sigma}{d\phi} = -1-\frac{r}{n_O}\frac{dn_O}{d r},
\end{eqnarray}
that is valid insofar as the following condition is satisfied
\begin{equation}\label{raapp}
    \frac{r_{H,obj}}{d}<\frac{r_{H,obj}}{r_{obj}}\ll1,
\end{equation}
where we assumed $r_{H,obj}$ and $r_{obj}$ as the horizon and size of the given compact object that generates the gravitational field associated with $n_O$, with $d$ the distance between the observer, supposedly the Earth, and the compact object itself.

Accordingly, Eq. \eqref{eq:9}, using the Schwarzschild metric, fulfilling the condition in Eq. \eqref{raapp},  can be rewritten as
\begin{equation}\label{scheq}
    \frac{d\sigma}{d\phi}\Big|_S=-\frac12 \left(1 - \frac{2r_H}{r-r_H} + \frac{r}{r - r_H \sin^2\sigma}\right),
\end{equation}
where the subscript, $S$, indicates the Schwarzschild case, with $r_H=2M$, as above.

Quite remarkably, note that Eq. \eqref{eq:9} remains unaltered under the discrete symmetry $n_O\rightarrow-n_O$, indicating that light deflection behaves the same for both spacetime material and gravitational metamaterial behaviors. This outcome is compatible with Eq. \eqref{intermedia}, where the switch $n_O\rightarrow-n_O$ has no impact on $\Phi$.

In addition, Eq. \eqref{scheq} exhibits a term $\propto \sin^2\sigma$. For small deviations this term is clearly negligible and, so, the maximum and minimum values associated with this expression are due to the choices $\sigma=0$ and $\sigma=\frac{\pi}{2}$, \textit{i.e.}, the two limiting cases of non-deflection and maximal deflection. Below we will consider such two cases to point out the extremes while computing $\sigma^\prime$.

Afterwards, ensuring $Q,q$ and $\Lambda$ to be much smaller than the mass, generating the compact object under exam, we argue the departures from the Schwarzschild case for Reisser-Nordstr${\rm \ddot{o}}$m, Bardeen, Schwarzschild-de Sitter and the AdS, quasi-anti de Sitter metrics, splitting the cases into:

\begin{widetext}

\begin{itemize}
    \item[-] \emph{Black holes}:
\begin{subequations}
    \begin{align}
        \frac{d\sigma}{d\phi}\Big|_{RN}&= \frac{d\sigma}{d\phi}\Big|_S-\frac{n Q^2 (2r-r_H)}{2r (r-r_H)^2},\\
        \frac{d\sigma}{d\phi}\Big|_{SdS}&=\frac{d\sigma}{d\phi}\Big|_S+\frac{n\Lambda r^3(2r-3r_H)}{6(r-r_H)^2},
    \end{align}
\end{subequations}
with $n=1$ or $n=2$ if $\sigma=\frac{\pi}{2}$ or $\sigma=0$, for the second and first orders around small $Q$ and $\Lambda$, respectively.
\item[-] \emph{Regular black holes}:
\begin{subequations}
    \begin{align}
        &\frac{d\sigma}{d\phi}\Big|_{B}=\frac{d\sigma}{d\phi}\Big|_S-\frac{3 M q^2 (3 r-2r_H)}{n r^2 (r-r_H)^2},\\
        &\frac{d\sigma}{d\phi}\Big|_{H}=\frac{d\sigma}{d\phi}\Big|_S-\frac{na^2r_H(4r-3r_H)}{r^3(r-r_H)^2},\\
        &\frac{d\sigma}{d\phi}\Big|_{D}=\frac{d\sigma}{d\phi}\Big|_S+{\small \frac{(4r_H L_D)^{2 - n}(2 L_D r_H ((n + 2) r_H - 2n^2 r) + (2 - n) \pi r (\frac{r_H^2}{2} - \frac{3 r_H}{2} r + r^2))}{n (r_H - r) (4r_H L_D  + \pi r (r_H - r))^{3 - n}}},\\
        &\frac{d\sigma}{d\phi}\Big|_{FW}=\frac{d\sigma}{d\phi}\Big|_S-\frac{3l_{FW}(2r-r_H)}{nr(r-r_H)^2},
    \end{align}
\end{subequations}
where, in the case of the Hayward metric,  the expansion is performed in terms of small ratio $a^2\equiv\frac{M}{\Lambda}$.
\item[-] \emph{Solutions with no horizons}:
\begin{subequations}
    \begin{align}
    \frac{d\sigma}{d\phi}\Big|_{aS}&=\frac{d\sigma}{d\phi}\Big|_S-\frac{nr_H}{r},\\
        \frac{d\sigma}{d\phi}\Big|_{AdS}&=\frac{d\sigma}{d\phi}\Big|_S+\frac{n}{3}\left(|\Lambda|r^2-\frac{3r_H}{r-r_H}\right),\\
        \frac{d\sigma}{d\phi}\Big|_{quasi-AdS}&=\frac{d\sigma}{d\phi}\Big|_S
        +(n - 1) - \frac{r_H}{r-r_H}+\frac{1 - n}{
 1 + k_0r^2} + (2 - n) \frac{|\Lambda|}{3}r^2.
    \end{align}
\end{subequations}
where {\rm $aS$} stands for the case of Schwarzschild for negative masses.
\end{itemize}

\end{widetext}

In the case of black holes, the effects due to $Q$ correspond to a second order correction, whereas the effects due to $\Lambda$ shifts $\sigma^\prime$ at first order.

Overall, the action of $\Lambda$ acts, therefore, more significantly than a charge. The same happens for regular black holes and for solutions without horizons.

\subsection{Distinguishing regular black holes from black holes}

The effects of a gravitational field can be, in principle, classified with respect to the source generating them. In this section we distinguish the effects due to black holes from those imagined from regular black holes. To do so, we categorize our analysis into:

\begin{itemize}
    \item[-] \emph{Weak gravitational field}. This case is identified by small stars, for example. In this respect, we focus on the Sun, which offers the great advantage of permitting local observations. We thus  employ \cite{2021LRSP...18....2C}

\begin{subequations}
\begin{align}
&M_\odot = (1.988475\pm0.000092)\times 10^{30}\,kg\,,\\
&r_\odot = (696192 \pm 247)\times 10^3\,m\,,\\
&r_{H,\odot}=(2951.440\pm0.137)\,m,
\end{align}
\end{subequations}
where
\begin{equation}\label{validity}
r_{H,\odot}/r_\odot\simeq 4.24\times10^{-6}\ll1,
\end{equation}
guaranteeing that the approximation, in Eq. \eqref{raapp}, holds for a ray passing close to the Sun within  small angles, namely under the  \emph{radial gravity approximation}.

Hence, by virtue of Eq. \eqref{nS}, after manipulation, Eq. \eqref{scheq} becomes
\begin{eqnarray}\label{eq:11}
\alpha^\prime_{S}=-1 + \frac{r_{H,\odot}}{r - r_{H,\odot}}\simeq -1+\frac{r_{H,\odot}}{r_\odot}\cos\phi,
\end{eqnarray}
where to a very good approximation, we consider

\begin{align}
 &r - r_{H,\odot}\simeq r_\odot/\cos(\phi).
\end{align}

The effect of gravitational deflection implies a departure, given by the integration of Eq. \eqref{eq:11},  yielding,
\begin{eqnarray}\label{eq:12}
\int_{-{\pi\over2}}^{{\pi\over2}}\frac{\cos\phi r_{H,\odot}}{r_\odot}\,d\phi\simeq\frac{2r_{H,\odot}}{r_\odot},
\end{eqnarray}
that provides
\begin{equation}
    \alpha^\prime_{S}\simeq (1.7489\pm0.0007)\mbox{ arc sec},
\end{equation}

confirming the consistency between the results experimentally found about light deflection by the Sun and our predictions.

Indeed, in the context of standard lensing, one can find the deflection angle, $\theta_{def}$, given by \cite{Cunha:2018acu}
\begin{equation}
    \theta_{def}=\frac{4M}{b},
\end{equation}
where $b$ is  the usual impact parameter. The above formula is equivalent to Eq. \eqref{eq:12}, determined in an alternative way, making use of the refractive index, $n_O$ and confirming the goodness of our approach.

On the other side, inserting nonzero $Q,\Lambda$ or $q$, appears unsuitable for the regime of weak gravity. Stars, like the Sun, may exhibit an intrinsic charge, typically $\sim 77C$, while around the system, as due to electron emission, we may have a significant contribution given by the net Sun charge roughly lying on $Q\sim 10^9 C$ \cite{2025AcAau.226..555P}. However, this  correction would just
imply a shift in $\alpha^\prime$ of about one part over $\sim10^{43}$, \textit{i.e.}, much smaller than any detectable effect. Hence, the corrections due to additional parameters, such as charge or vacuum energy, appear impossible to be measured in regimes of so small gravity.

\item[-] \emph{Strong gravitational field}. It is identified by dense stars, for example. In this respect, we focus on neutron stars and white dwarfs, assuming for them two maximally-rotating configurations. Here, we do expect larger deviations than the results previously obtained for the Sun.

So, considering the specific maximally-rotating configurations, labeled \emph{NS} and \emph{WD}, for neutron stars and white dwarfs, respectively \cite{Boshkayev:2020igc},

\begin{subequations}\label{conditionsNS}
    \begin{align}
        M_{NS}&\in[1.00;2.18]M_\odot,\\
        r_{NS}&\in[10;15] km,\\
        M_{WD}&\in[0.15;1.33]M_\odot,\\
        r_{WD}&\in[4;18]\times 10^3 km,
    \end{align}
\end{subequations}
we can infer the magnitude of light deviation, following the same strategy pursued for the Sun. Thus, we immediately generalize Eqs. \eqref{scheq} and \eqref{eq:11} by
\begin{subequations}\label{NSdev}
    \begin{align}
    \alpha^\prime_{S,NS}\simeq -1+\frac{r_{H,NS}}{r_{NS}}\cos\phi,\\
    \alpha^\prime_{S,WD}\simeq -1+\frac{r_{H,WD}}{r_{WD}}\cos\phi,
\end{align}
\end{subequations}

and, then, by virtue of Eqs. \eqref{conditionsNS}, with the usual prescription $r_{H,\{NS,WD\}}=2M_{\{NS,WD\}}$, Eqs. \eqref{NSdev} yield
\begin{subequations}\label{nsewd}
    \begin{align}
    &\alpha^\prime_{S,NS}\in [0.39;1.29]\mbox{ rad},\\
    &\alpha^\prime_{S,WD}\in [10.15;404.84]\mbox{ arc seconds},
    \end{align}
\end{subequations}

that have been computed using the maximum and minimum values that can be obtained substituting the values of mass and radii inside the expression, above considered. In particular, we obtained the maxima with the \emph{largest masses and smallest radii}.

Hence, the overall magnitudes are set to be around the unit of radiant for neutron stars and roughly houndreds of arcsecs for white dwarfs.

Thus, we establish nonzero contributions for the charges, $Q$ and $q$, and the cosmological constant, $\Lambda$, as reported in Table \ref{due}. Those findings have been found considering the deviations determined in the above relations and open avenues to measure quantities such as charges or de Sitter phases, by means of optical properties. As a consequences, there is no evidence to discard the use of regular solutions that may be used to characterize such objects, alternatively to the Schwarzschild solution. Conversely, as due to the degeneracy between regular spacetimes, it appears unclear how to handle alternatives to Bardeen and Hayward metrics, suggesting that the Dymnikova and Fang-Wang spacetimes are unlikely to describe neutron stars or white dwarfs in view of a non-clear interpretation for $l_{FW}$ and $l_D$.

\item[-] \emph{Distinguishing solutions different from black holes?}

Approaching the horizons, the refractive index becomes infinite for both black holes and regular black holes. Accordingly, the effects due to light deflection become particularly intense, as expected. To detect objects different from solutions with horizons, namely naked singularities or wormholes, one has to check whether light modifies the deflection smoothly while pointing out the singularity. In such a case, it would be possible to distinguish between black holes and naked singularity.

The same cannot happen between black holes and regular black holes since, to distinguish their presence from black holes, one would need to check the behavior \emph{inside the horizon} and this turns out to be unpracticable. However, imaging to have information about the masses  around such solutions, computing the refraction at different radii, it is possible to speculate if a compact object is a black hole or another kinds of mimicker, such as a no-horizon regular or naked metric. In principle, this would be feasible due to the evident $n_O$ magnitudes, across the three classes of solutions, as prompted in Figs.~\ref{deltaindiciperBHs}, \ref{indiciperRBHs} and \ref{antideSitter}.

However, one can theoretically compute $n_O$ for very intense gravitational fields, violating \emph{de facto} Eq. \eqref{interm}, and re-evaluate the deflection angles that, once detected, can help in classifying future likely observed exotic objects, such as strange stars, boson, quark stars, and so on.

\end{itemize}

\begin{table*}
\setlength{\tabcolsep}{1.0em}
\renewcommand{\arraystretch}{1.2}
\caption{Principal parameters and results involved in our pictures of weak and strong gravitational fields. The table summarizes forecasts related to the additional quantities that enter the underlying spacetimes.}

\begin{tabular}{lcl}
\hline\hline
\multicolumn{3}{c}{{\bf Regime of weak gravity}}\\
\cline{1-3}
		& $Main\,\,\,parameters$ & $Outstanding\,\,\,results$ \\

\hline
		&  $M\in[M_\oplus;M_\odot]$ & Results using different metrics are indistinguishable.\\

{\bf \emph{Solar System}}		&  $Q$  & Spherical symmetric effects are uniquely measured.\\

		& $\Lambda$ & No vacuum contribution detectable.\\
        & $q$ & Topological charge smaller than physical charge.\\
& $L_D, l_{fw}$ & Non-physical quantities describing the Sun whose effects are negligible.\\

\hline\hline
\multicolumn{3}{c}{{\bf Regime of strong gravity}}\\
\cline{1-3}
		& $Main\,\,\,parameters$ & $Outstanding\,\,\,results$ \\

\hline
		& $M\in[0.15;2.18]\,M_\odot$ & Deviations are observable only for  maximally-rotating compact objects.\\

{\bf \emph{NSs} $\,\&\,$ \emph{WDs}}		&      $Q_{NS}\simeq [1;1.6]\times 10^{19}\,C$
& The charge influences roughly of $5\%$ on the the overall outcome.\\
		&
         $\Lambda_{NS}\leq (4.76\div9.53)\times 10^{-17}m^{-2}$ & Instrument sensibility  can probe $1\sigma$ deviations from spherical symmetry. \\
	& $\Lambda_{WD}$ and $Q_{WD}$ &  Both smaller than those predicted for neutron stars of a factor 10.\\

\hline\hline
\end{tabular}

\label{due}
\end{table*}

\section{Refractive index in flat spacetime}\label{sezione6}

As before reported, a generic curved spacetime generates   inhomogenous and isotropic contributions to $n_O$. Focusing on flat space, it is therefore natural to question whether flatness naturally leads to $n_O=1$.

Immediately a flat Minkowski metric, in spherical coordinates, \textit{i.e.}, $ds^2=-dt^2+dr^2+r^2d\Omega^2$, provides $n_O=1$, implying that a flat space in spherical coordinates leads to a \emph{perfect lens with positive refractive index}. In this case, a metamaterial-like behavior of the spacetime medium is unlikely, since the speed of light keeps  positive sign.

With the aim at generalizing, focus on the Rindler coordinates. So, starting from $ds^2=-dT^2+dX^2+dY^2+dZ^2$ and providing the following transformations,
\begin{subequations}\label{Rindlerchoice}
    \begin{align}
        &X=\sinh{(a t)},\\
        &T=\cosh{(a t)},\\
        &Y=y,\quad        Z=z,
    \end{align}
\end{subequations}
in Cartesian coordinates, we immediately obtain \cite{Balasubramanian:2013rqa}
\begin{equation}\label{Rindler1}
    ds^2=-(a x)^2dt^2+dx^2+dy^2+dz^2.
\end{equation}
The parameter $a$  represents the proper acceleration, defined along the hyperbola $x=\frac{1}{a}$. The choice of Rindler coordinates, in Eqs. \eqref{Rindlerchoice}, appears of particular interest as they are related to the Unruh effect \cite{Crispino:2007eb}, providing the existence of a Unruh temperature, in principle measurable and associated with the free parameter, $a$.

Consider now the way to extend Eq. \eqref{Rindler1} to spherical coordinates. Hence, starting from a generic Minkowski space in spherical coordinates, $ds^2=-d\eta^2+d\rho^2+\rho^2d\Omega^2$ and adopting the transformations,
\begin{subequations}\label{trasformazioniRindler}
    \begin{align}
        &r=\sqrt{(\rho-b)^2-\eta^2},\\
        &At=\tanh^{-1}\left(\frac{\eta}{\rho-b}\right),\label{AT}
    \end{align}
\end{subequations}
where $\rho>0$ and $\eta>0$, we end up with
\begin{equation}\label{Rindler2}
    ds^2=-A^2r^2dt^2+dr^2+r^2\left(\frac{b}{r}+\cosh\left(At\right)\right)^2d\Omega^2,
\end{equation}
that turns out to represent the Rindler spacetime in spherical coordinates, particularly useful to  compute the refractive indexes. Here,  $b>0$ and, remarkably, a static observer in spherical Rindler frame at constant $r$ is moving hyperbolically as viewed from the Minkowski space. Conversely, other accelerating observers cannot see the
interior of the sphere, whose center is placed in the origin of Minkowski
coordinates, being here characterized by the radius $b$.

Hence, considering Eq. \eqref{indicedirifrazione1}, the refractive index in Rindler coordinates reads,
\begin{equation}\label{nOperRindler}
    n_O^{R}=\frac{1}{Ar}\,\sqrt{1-\sin^2\sigma\left[1-\left(\frac{b}{r} + \cosh(At)\right)^2\right]},
\end{equation}
that turns out to be positive (negative) when $A>0$ ($A<0$).

At a first glance, since $A$ is associated with the Unruh radiation, a negative value can lead to negative Unruh temperature that appears unphysical or, in general, hard to interpret \cite{Volovik:2023stf}.

However, in determining the Unruh temperature, the $A$ sign appears irrelevant, as the temperature depends on its modulus by
\begin{equation}\label{fav}
    T_U=\frac{|A|}{2\pi k_B},
\end{equation}
where $k_B$ is the Boltzmann constant. The above definition suggests two main consequences. First, Eq. \eqref{nOperRindler} defines a way to directly measure the refractive index for an accelerated observer, adopting the Unruh temperature. In other words, measuring the Unruh temperature implies a measure of the $n_O$ magnitude. Second, from Eq. \eqref{nOperRindler}, $n_O\neq1$ even in flat space, being, moreover, $r$-dependent.

Further, to understand the physical meaning of a negative $A$, it is useful to examine the Ricci scalar, evaluated for Eq. \eqref{Rindler2}
\begin{equation}
    R=\frac{(A^2-1)(r + 4 b \cosh t +
   3 r \cosh 2t)}{A^2 r (b + r \cosh t)^2},
\end{equation}
that turns out to vanish only if $A=\pm1$, resembling the case of flat space as in Cartesian coordinates, where the curvature vanishes. Again, this indicates either a positive ($A=+1$) or negative ($A=-1$), albeit $r$-dependent, refraction even in flat space.

Invoking that the spherically-symmetric version of the Rindler coordinates is analogous to the Cartesian case, by virtue of Eq. \eqref{AT}, it appears evident that, in order to have $A<0$, one has to lie on $b>\rho$, namely the observer would be placed \emph{inside the sphere that we used to construct the metric itself}. This situation is similar to the sign switch of $n_O$ inside the horizon for black holes. Alternatively, if $A<0$, then to guarantee that $\rho-b>0$, one needs $t<0$, giving rise, in any cases, to a positive Unruh temperature, by virtue of Eq. \eqref{fav}.

Summarizing the above results, we can end up with the following physical outcomes.
\begin{itemize}
    \item[-]  The effects of curvature \emph{are not directly} responsible for having a metamaterial behavior of spacetime, in fact, with zero curvature it is possible to have $|n_O|\geq1$, as it happens in Schwarzschild for example, as well as in the Rindler case, above described.
    \item[-] A flat space is a necessary but not a sufficient condition to have a perfect lens behavior, or alternatively a homogeneous refractive index, as shown for the Rindler coordinates.
    \item[-] A negative refractive index can be pursued inside the sphere to construct the metric,  similarly to the Schwarzschild case. Alternatively, inverting the sign of $t$ can lead to an analogous result, without limiting to be inside  the Rindler sphere.
\end{itemize}

We can now wonder whether gravitational metamaterials may lead to any possible physical interpretation associated with compact objects that exhibit such properties.

Motivated by these reasons, in the next section, we propose and investigate their possible (classical) particle-like behavior, while later on, we will turn to their possible quantum nature.

\section{Spacetime as gravitational metamaterial}\label{sezione7}

Experimentally, it is possible to use materials to manipulate the electromagnetic fields \cite{metamatbook} and, consequently, to exploit their optical properties to behave exotically.

Naively, if the materials are homogeneous, the interface between them can be strategically utilized to guide electromagnetic fields in a desired manner. Hence, in solid state physics, metamaterials, as novel classes of materials, have reached great attention during recent years, as above explained.

For example, handling a material that exhibits \emph{simultaneously} an effective negative electric and magnetic permeability may lead to a negative refractive index\footnote{Traditionally, metamaterials are made of metallic
nanostructures, arranged periodically, whose size appears much smaller than the incident wavelength, exhibiting a bulk optical response. Nevertheless, metamaterials have been constructed as artificially engineered materials with
exotic optical properties as pioneering conjectured in 1967 by Vaselego \cite{vaselago}.} \cite{metasperimentali}.

Above, we proposed that, as consequence of general relativity, a given curved spacetime may exhibit properties that resemble those of metamaterials, since we predicted the existence of negative refractive index, working on the sign of lapse and/or shift functions.

Accordingly, one can argue if additional properties of such materials may be recovered in gravitational contexts and if there are fields of application for precise Einstein's gravity solutions.

For example, in Ref. \cite{Pendry} was shown how to obtain a perfect lens, opening new avenues toward the intriguing property of invisibility \cite{Fleury}.

Without entering deeply the subject of solid state physics metamaterials, we here propose that it is possible to infer  gravitational metamaterials if

\begin{itemize}
    \item[-] the spacetime is modeled in terms of a medium, exhibiting, moreover, a negative refractive index;
    \item[-] it shows properties that resemble or mime those of a particle\footnote{For the sake of completeness, gravitational metamaterials are hereafter a proposal to explain that spacetime configurations can be used to both model media, but also to characterize optical properties in an unexpected and unexplored way. The spacetime properties in view of negative refractive indexes are, in fact, described below and used to describe possible particle-like configurations mimicking, for example, particles of dark matter. }.
\end{itemize}

Consequently, can given spacetimes mime particles while remaining invisible? To characterize this, we examine whether metrics with a negative refractive index can show particle-like behavior, drawing analogy with the interpretation of \emph{spacetime as particles} \cite{Holzhey:1991bx}.

To do so, we speculate on toy models that resemble the behavior of such solutions. We then point out a possible interpretation suggesting quantum properties of these particle-like configurations \cite{tHooft:1984kcu}. Accordingly, we reinterpret these configurations as quasiparticles, as recently proposed in e.g., Refs. \cite{Belfiglio:2022cnd,Belfiglio:2022qai,Belfiglio:2024swy,Luongo:2023aaq}.

\subsection{Particle-like behavior of gravitational metamaterials}

A standard spacetime solution can exhibit properties that resemble those of a particle, under certain circumstances. This topic has been faced in Ref. \cite{1997IJTP...36.1475R} and generalized for black holes in Ref. \cite{Holzhey:1991bx}. Hence, the conditions that we propose  for gravitational metamaterials to behave particle-like are listed below.
\begin{itemize}
    \item[i)] The spacetime might be regular and stationary.
    \item[ii)] The solution might at least recover the Schwarzschild metric at large radii or, more broadly, flatter at infinity.
    \item[iii)] The energy density and pressures might be finite. Toward the pressures, it is not possible to make any definitive statement \emph{a priori}. One can only speculate about the conditions under which the particle-like setting could behave like dust, for example looking at the nature of the corresponding equation of state.
    \item[iv)] The solution might not have thermodynamic properties such as entropy, temperature, etc.
    \item[v)] The solution might show precise optical properties that, under precise conditions, may even predict negative refractive indexes.
\end{itemize}

Particularly, let us focus on the fourth and fifth conditions. For example, in the Reissner-Nordstr$\ddot {\rm o}$m solution, the extreme black hole,
$M=Q$, implies zero temperature, albeit the entropy, proportional to the horizon area, still does not vanish. To ensure that the above properties provide a particle-like behavior, the corresponding solutions might not exhibit thermodynamic properties, so neither temperature nor entropy. Generalizing, potential examples, satisfying the five requirements simultaneously, cannot be constructed \emph{ad hoc} with two coinciding horizons, since they may still lead to a nonzero entropy contribution, despite a vanishing temperature.

Likely the simplest possibility, that involves the temperature and entropy to vanish, may occur in the absence of horizons, e.g., in configurations such as naked singularities or wormholes. However, since the full set of conditions excludes the presence of singularities, we conclude that the most viable spacetimes possibly turn into \emph{no-horizon regular solutions}.

Below, we present three toy models that satisfy the five conditions, required above, to show a particle-like behavior.

\section{Examples of gravitational metamaterials}\label{sezione8}

We previously showed that, even in flat space, the refractive index may turn into negative values. Thus, for practical cases, negative refraction can be displayed by
\begin{itemize}
    \item[-] homogeneous $n_O$, that are characterized by a constant $n_O$;
    \item[-] inhomogeneous $n_O$, where $n_O=n_O(r)$.
\end{itemize}

As plausible toy models, considering the above five conditions, we distinguish:
\begin{itemize}
    \item[-] Solutions with homogeneous $n_O$.
    \item[-] Solutions with weak gravitational fields and inhomogeneous $n_O$.
    \item[-] Regularized solutions, constructed from black holes, exhibiting no horizons.
\end{itemize}

The three approaches will be described simplifying conventionally $n_O$, employing $\sigma=0$ and $\gamma=1$.

\subsection{Homogeneous gravitational metamaterials: Conformal metric}

In the case of isotropic spacetime media, homogeneous gravitational metamaterials can be conventionally obtained requiring $n_O=\mu$, with $\mu$ arbitrary negative constant. Without losing generality, one can assume that $\mu=-1$, here implying
\begin{equation}\label{gm1}
    ds^2=-\alpha^2(dt^2-dr^2)-r^2d\Omega^2,
\end{equation}
turning out to be equivalent to a \emph{conformal metric}, Eq. \eqref{primaconforme}, say
\begin{equation}\label{gm2}
    ds^2=-\alpha^2(dt^2-dr^2-\mathcal R^2d\Omega^2),
\end{equation}
where $\mathcal{R}\equiv \frac{r^2}{\alpha^2}$, namely \emph{to have an homogeneous gravitational metamaterial is to have a conformally flat spacetime}, meanwhile ensuring  $\mu<0$.

Using Eq. \eqref{gm1}, it is useful to compute the  Einstein equations,
\begin{subequations}\label{equazionidicampo}
    \begin{align}
        \rho&=-\frac{\alpha - \alpha^3 - 2r \alpha^\prime}{r^2 \alpha^3},\\
        P_r&=\frac{\alpha - \alpha^3 + 2r \alpha^\prime}{r^2 \alpha^3},\\
        P_t&=-\frac{\alpha^{\prime\,2} - \alpha \alpha^{\prime\prime}}{  \alpha^4},
    \end{align}
\end{subequations}
where $\rho$ is the energy density and $P_r$, $P_t$ are the radial and tangential pressures, respectively.

To compute $\alpha$ through Eqs. \eqref{equazionidicampo}, we require a further constraint, under the form of an the equation of state.

Excluding the trivial case of constant density, the possibly simplest  assumption lies on assuming that  one pressure, either radial or tangential, turns out to be constant. Accordingly, we can therefore assume
\begin{align}
P_r&=P_{r,0},\label{prcostante}
\end{align}
where $P_{r,0}$ is the value of the radial pressure at all radii.

Hence, we obtain
\begin{align}\label{alphaquasip1}
&\alpha=\pm\left[1-\left(\sqrt{P_{r,0}}r - {\mathcal K\over 2 \sqrt{P_{r,0}}}\right)^2 + {\mathcal K^2\over 4 P_{r,0}}\right]^{-1},
\end{align}
where the negative sign is extracted to guarantee the fifth condition on gravitational metamaterials. So, considering the associated mass,          $m\equiv\int \rho\,r^2dr$, \textit{i.e.}, the mass contained inside spherical symmetric halo\footnote{In spherical symmetry, since we are dealing with compact objects characterized by spacetimes, the MS definition in Eq. \eqref{MSdefinition}, essentially coincides with the constituent mass, here considered. }, we immediately obtain
\begin{subequations}
\begin{align}
    \rho&=3 P_{r,0} - \frac{2 \mathcal K}{r},\\
    m&=r^2 (P_{r,0} r-\mathcal K),\\
    R&=4 P_{r,0} - {2 \mathcal K\over r} - \frac{\mathcal K^2 + 4 P_{r,0}}{1 + \mathcal K r - P_{r,0} r^2}.
\end{align}
\end{subequations}
To guarantee that the density and mass are non-singular and finite, with no horizons, we fix the integration constant, $\mathcal K$, and the radial pressure, $P_{r,0}$, to
\begin{equation}\label{conditiosuperconstants}
    \mathcal K=0\,,\quad P_{r,0}<0.
\end{equation}

This first attempt to obtain a constant refractive index, namely an homogeneous medium behavior, with a particle-like behavior is thus characterized by
\begin{itemize}
    \item[-] \emph{advantages}: The density and pressures are finite and at infinity $|\alpha|\rightarrow1$. If $P_{r,0}$ is small enough, it can mime a fluid that has a constant, but weak, density and radial pressure, different from a pure de Sitter solution (since $P_{t}$ is not a constant). The solution is regular and exhibits no horizons, having negative refraction if $\mu<0$.
    \item[-] \emph{disadvantages}: The strong energy condition is violated, as well as the mass and density of particle-like contribution, which are negative definite\footnote{ For example, ensuring that the constituent is dark matter, the study of  negative massive particles is a possibility already explored, see e.g.,  \cite{Socas-Navarro:2019pps,Farnes:2017gbf}, albeit this may lead to  exotic scenarios, not fully understood so far.}. Moreover, the equation of state for the radial pressure reads $
    \omega_r\equiv \frac{P_{r,0}}{\rho}=\frac{1}{3}$. This shows that the fluid closely resembles cosmological radiation, indicating that the radial pressure magnitude is comparable with $\rho$.
\end{itemize}

More complicated approaches may be explored by assuming \emph{a priori} that $\omega_r$ is fixed and small enough. Hence, a more viable and refined solution would require to infer the correct equation of state, formulated in terms of the specific fluid constituent under exam, \textit{i.e.}, an aspect that lies beyond the scope of the present work.

\subsection{Inhomogeneous gravitational metamaterials: Newtonian metric}

There is a \emph{plethora} of spacetimes that may lead to inhomogeneous refractive indexes. Conventionally, we here single out likely the simplest spacetime, provided by the Newtonian metric,
\begin{equation}\label{metricanewtoniana}
    ds^2=-(1+2\Psi_1)dt^2+(1-2\Psi_2)dr^2+r^2d\Omega^2,
\end{equation}
hereafter employed in the Newtonian gauge, \textit{i.e.}, $\Psi_1=\Psi_2=\Psi$, for the sake of simplicity.

The Newtonian metric \emph{is neither} regular \emph{nor} horizon-less, in general. However, we assume here that $|\Psi|\ll \frac{1}{2}$, to ensure that there are no horizons, requiring in addition global regularity. This metric can therefore be seen as a very simple prototype to give an inhomogeneous refractive index, involving a very weak gravitational field.

Following the recipe of Eq. \eqref{prcostante}, the potential then reads
    \begin{equation}\label{equazioneperpsi}
        \Psi=
        \frac12 - \left[1 + \frac{2 e^{P_{r,0} r^2} \sqrt{P_{r,0}} r}        {
 4 k\sqrt{P_{r,0}}  + \sqrt{\pi} \mathcal E(\sqrt{P_{r,0} r})} \right]^{-1},
    \end{equation}
where $k$ is an integration constant and $\mathcal{E}\equiv {\rm erf}(iz)/i$, involving the error function, ${\rm erf}\equiv \frac{2}{\sqrt{\pi}}\int_0^z e^{-t^2}dt$.

Accordingly, the Einstein equations and associated constituent mass become
\begin{subequations}
    \begin{align}
        \rho&\simeq -P_{r,0} - \frac{4 P_{r,0}^2 r^2}{9},\\
        P_t&\simeq P_{r,0},\\
        m&\simeq -\frac{4\pi}{3} P_{r,0}r^3 \left(1 + \frac{4 P_{r,0} r^2}{15}\right),
    \end{align}
\end{subequations}
where we ensured that $P_r=P_{r,0} $, by expanding the density, tangential pressure and matter up to the second order in $|P_{r,0}|\ll1$, with arbitrarily vanishing $k$ and negative $P_{r,0}$. This scheme turns out to be a good approximation for two main reasons. First, the pressure is taken to be small in order to guarantee that the constituent behaves similarly to a genuine pressureless fluid. Second, $k=0$ also avoids singularities in the density.

With this prescription, we also observe an increase of mass and a quite stable tangential pressure, constant at $r=0$, and negative definite, for $P_{r,0}<0$. In particular, at $r=0$, we have
\begin{align}
    &\rho(0)=-P_{r,0},\\
    &P_{t}(0)=P_{r,0},\\
    &R(0)=-2P_{r,0},
\end{align}
there, quite interestingly the core equation of state for radial pressure reads,
\begin{equation}
    w_{r}(0)\equiv\frac{P_{r,0}}{\rho(0)}=-1,
\end{equation}
resembling the properties of a cosmological constant\footnote{Asymptotic behaviors of spacetime, resembling the behavior of quintessence and/or cosmological constant, have been discussed in Ref. \cite{miametrica}.}.

Even though we imposed regularity, the underlying metric also requires to be free of horizons. This prerogative has been ensured by setting the gravitational potentials to remain very small.

However, to guarantee that our second order mass approximation holds, an observer can be placed before the radius where $m$ vanishes, say $r_c$.

In particular, $m=0$ implies $
r_c\simeq\frac{1}{2}\sqrt{\frac{15}{|P_{r,0}|}}$, up to the second order of Taylor expansion for very small $P_{r,0}$.

Hence, by restricting to $r\ll r_c$ and using Eq.~\eqref{equazioneperpsi}, this critical radius, $r_c$,  can be compared with galactic structures. For example, $P_{r,0}\simeq -10^{-5}$ implies $r_c\simeq 120$, with $|\Psi|\lesssim 2.5\cdot 10^{-2}$ at $r=\frac{r_c}{5}$.

The second attempt to obtain an inhomogeneous refractive index, with a particle-like behavior is somehow more promising than the first and characterized by
\begin{itemize}
    \item[-] \emph{advantages}: The radial pressure is again constant and small, mimicking a fluid with proper density and negative tangential pressure, if $P_{r,0}<0$, in analogy to dark energy contexts \cite{Copeland:2006wr}. Thus, the solution is regular and the mass appears positive definite. Finally, negative refraction may occur imposing $\alpha=-\sqrt{1+2\Psi}$.
    \item[-] \emph{disadvantages}: The model provides negative pressures that in modulus appear comparable with the density magnitude. This suggests a fluid constituent with negative equation of state\footnote{A negative equation of state for dark matter is also found in Ref. \cite{Faber:2005xc}.}. Moreover, the metric can describe this fluid configuration within a precise interval of radii, where the potential remains small enough.
\end{itemize}

\subsection{Regularized spacetimes without horizon}

Compact objects without horizons are not particularly easy to justify in general relativity. Black holes or regular black holes exhibit naturally horizons, for example, while assuming objects that, instead, do not show horizons is generally limited to naked singularities and wormholes.

Discarding the very peculiar role played by wormholes, focusing on naked singularities, one can notice that a very immediate example of naked singularity is offered by the Reisser-Nordstrom spacetime with $Q>M$, as discussed previously. For the sake of completeness, in general, there exist several options to obtain, from a given metric, the corresponding horizon-free counterpart. Naked singularities, in general, exhibit divergences at $r=0$, and there are speculations toward their existence, as consequences of the cosmic censorship conjecture  \cite{Lehner:2010pn,Figueras:2017zwa}.

From the treatment followed in the previous two cases and fulfilling all the above five conditions, we require a \emph{no-horizon regular solution}. Nevertheless, a prototype of such a solution has been prompted above. The Newtonian metric, in fact, guaranteeing that $\Psi\ll{1\over2}$ and the mass is limited within the interval that goes up to the critical radius, allows to formulate a horizon-free metric that is also regular at the same time, if $P_r=P_{r,0}$.

It is natural to ask whether the construction of regular, horizon-free spacetimes can be extended to a broader class of solutions.

To do so, a singularity-free spacetime can be obtained through the phenomenological procedure of regularization proposed by Simpson and Visser \cite{Simpson:2018tsi} that, considering an auxiliary new variable, $u$, interestingly defines
\begin{equation}\label{regolarizzazione}
r(u) = \sqrt{u^2 + B^2},
\end{equation}
where $B$ is the regularization parameter, \textit{i.e.}, the singularity at $r=0$ is replaced by a regular minimum of Eq. \eqref{regolarizzazione}. Ensuring no horizon yields\footnote{Here, we limit our analysis to two free parameters only, \textit{i.e.}, $M$ and $B$ only. Further parameters, such as charge or de Sitter phase would imply a different construction that can avoid the horizon.}
\begin{subequations}
    \begin{align}
\alpha&=\beta^{-1}=\pm\sqrt{1+\frac{2M}{\sqrt{u^2+B^2}}}\,,\label{S-V1}\\
\gamma&=\frac{\sqrt{r^2+B^2}}{r},\label{S-V2}
    \end{align}
\end{subequations}
where we regularize the Schwarzschild solution, using Eq. \eqref{SCHW},  switching the mass sign.

Further, at $u=0$, the spacetime corresponds to a sphere of radius $B$, whose physical interpretation is particularly relevant when the mass is positive. In this case, for  $B>2M$, the metric describes a wormhole with a throat at $u=0$ or a black hole with two horizons at  $u=\pm\sqrt{4M^2-B^2}$, if $B<2M$, and finally $B=2M$ implies an extremal black hole. In the latter case, however, the horizon exists and is placed at\footnote{In the case of a black hole, the hypersurface $u=0$ does not represent a throat, because $u$ is a temporal coordinate. The field sources responsible for the Simpson-Visser spacetimes were examined in Ref. \cite{Bronnikov:2021uta}.
} $u=0$.

Easily speaking, one can consider $M<0$ and $|B|>2|M|$ to avoid the presence of a wormhole, implying a regular solution without horizon.

Summing up, the structure of the solution has no horizon and non-singular, furnishing the solution that we are searching, as stated above.

Computing the density and pressure, we obtain
\begin{subequations}
    \begin{align}
&\rho=\frac{B^2 (4 M + \sqrt{B^2 + u^2})}{(B^2 + u^2)^{\frac{5}{2}}},\\
&P_r=-\frac{B^2}{(B^2 + u^2)^2},\\
&P_t=\frac{B^2 (M + \sqrt{B^2 + u^2})}{(B^2 + u^2)^{\frac{5}{2}}},
    \end{align}
\end{subequations}
and, accordingly, the MS mass
\begin{equation}
    m=\frac{2}{9} \pi
    \left[
    -\frac{8 M u^3}{(B^2 + u^2)^{\frac32}} - \frac{3 B^2 u}{B^2 + u^2} +
   3 B \arctan\left(\frac{u}{B}\right)
   \right].
\end{equation}

In this third example, again, an inhomogeneous refractive index can be computed, whereas a particle-like behavior is naively achieved and characterized by
\begin{itemize}
    \item[-] \emph{advantages}: The Simpson-Visser procedure can make metrics regular and they can match the five requirements above described;
    \item[-] \emph{disadvantages}: The mass is not completely positive-definite and the overall treatment remains a procedure \emph{ad hoc} to regularize metrics. Again, the pressure magnitudes are comparable to the density one.
\end{itemize}

The third example deals with a procedure that permits to retrieve regular solution adopting Eq. \eqref{regolarizzazione}. However, seeking no-horizon regular solutions implies additional free parameters, to guarantee simultaneously that the mass is not altered from its positive sign and the pressures can be negligible with respect to the density.

\subsection{Dark matter as gravitational metamaterial}

Describing spacetimes as particles-like  recalls the need of having fluids that can mime cold dark matter. Attributing to spacetimes the characteristics of particles and imaging that these objects may act as dark matter is a prerogative of this work and appears motivated by the absence of experimental results toward the discovery of dark matter particles under the form of weakly massive interacting particles (WIMPs). The WIMP paradigm has, in fact, encountered recent experimental incongruence shifting the  interest of the community to ultralight fields and, in particular, to axions \cite{Duffy:2009ig}.

However, a ``geometric" origin of dark matter is not a novel idea. Attempts to obtain dark matter in an alternative way involve extending gravity by extra terms \cite{Pal:2004ii}, modifying gravity at different scales \cite{Milgrom:2014usa} or including several new constituents, through effective theories going beyond the Standard Model of particle physics \cite{Levi:2018nxp}.

The approach of having alternatives \emph{inside} Einstein's theory, namely without extending the theory itself, is instead less explored. In this context, our proposal suggests that spacetimes behave as particles, but they are not strictly speaking particles, but more similar to solitons \cite{Brax:2025uaw}, geometric dark matter \cite{Demir:2020brg}, and/or quasiparticles \cite{Belfiglio:2022cnd,Belfiglio:2023moe,Belfiglio:2024swy,Belfiglio:2022cnd,Belfiglio:2024swy}.

As a matter of modern consensus, Newtonian physics appears quite mandatory in describing dark matter's nature, and this, together with our five conditions, stated above, leads to the following two main tips, \textit{i.e.},

\begin{itemize}
    \item[-] particle-like configurations from spacetime are possible dark matter candidates if the equations of state  remain small enough to guarantee Newtonian gravity to hold,
    \item[-] these candidates have to be compatible with the concept of being dark, namely no light emitted, absorbed, or reflected, making it directly unobservable and, in this respect, our fifth condition guarantees this prerogative.
\end{itemize}

Interestingly, the natural landscape in which one can immediately test the above toy models could be the context of spiral galaxies. There, observations require that,
\begin{enumerate}
\item the gravitational field is weak. For spirals, in particular, this appears evidently clear, apart from the central region;
\item probe particle speeds appear slower if compared to the speed of light, with the exception of gravitational lensing;
\item pressure magnitudes are extremely  small, \textit{i.e.}, smaller than density, as well as matter fluxes, if any, behaving analogously\footnote{In general, one can even expect that dark matter exhibits high pressures, but this hypothesis is the least likely, according to the actual consensus.}.
\end{enumerate}
So, addressing the first condition above, the gravitational potential, say $\Phi(r)$, fulfills the condition \citep{Misner:1973prb}
\begin{equation}\label{potenz}
\nabla^2 \Phi \approx R_{00},
\end{equation}
where the $00$ component of the Ricci tensor has been used, having $
R_{00}\sim -\rho + p_r + 2p_t$.

Thus, the TOV lapse function, $\Phi(r)$, can be therefore considered as the
Newtonian gravitational potential only when the pressures, regardless if radial or tangential, are negligible with respect to $\rho$.

Among the three cases discussed above, only the inhomogeneous model appears to be practically viable, as both the radial and tangential equations of state are negative definite. In addition, if the radial pressure is fixed to a small value, then $P_t$ is simultaneously small. Thus, it appears interesting to focus on this model and, therefore, from Eq. \eqref{potenz}, limiting to spirals, the velocity of a rotation curve can be approximated by
\begin{equation}\label{velocità}
    v_{tot}^2\simeq r\frac{d\Phi_{tot}}{dr}=v_{DM}^2+v_{N}^2,
\end{equation}
where the subscript ``${\rm N}$" stands for Newtonian, and therefore \emph{visible}, counterpart. Notice that $v_{DM}^2\equiv r\frac{d\Phi}{dr}$, while in Eq. \eqref{velocità} we neglected, for simplicity, the contributions due to the gas, disk, and so on. A more accurate analysis needs to consider such contributions, as they contribute, even significantly, to the net velocity, $v_{tot}$.

Accordingly, the velocities are induced only by gravitational potentials and, then, using indifferently either the TOV or Newtonian metric, up to the second order in $P_{r,0}$, we have
\begin{align}\label{nopriw}
    v_{DM}^{inh}&\simeq\frac{r}{3}\sqrt{\frac{2 P_{r,0}^2 r^2}{5}+3|P_{r,0}|}.
\end{align}
At this stage, it is remarkable to stress that spiral galaxy rotation curves \emph{are not flat} \cite{Mannheim:1996rv} but the velocities can increase, decrease, or remain roughly constant. This appears clear once computing velocities for given models, such as ours, as it turns out to be evident from Eq. \eqref{nopriw}. There, the asymptotic regime cannot be reached since the radii are clearly limited to the galaxy size, \textit{i.e.}, from the region far from the bulge up to the end of the galaxy itself. This picture is fully-compatible with the recipe found in our second example, described by an outer radius, where the MS mass vanishes.

In conclusion, there may be multiple solutions satisfying the five conditions outlined above, each corresponding to a specific velocity and, possibly, constructed using techniques, such as the regularization proposed as third example.

A direct comparison of Eq. \eqref{nopriw} with data and further explorations of alternatives to it, under the form of  no-horizon regular solutions of dark matter, satisfying the five conditions, go beyond the purposes of this work and will be objects of future speculations. Here, we provided the possibility to construct dark matter under the form of particle-like setups.

Quite remarkably, this treatment candidates as a compelling alternative to the search for dark matter in the form of beyond-Standard Model particles, but it appears purely classic, as \emph{it makes use of spacetimes providing negative refraction}.

However, in some cases, particle-like behavior can be associated with a quantum nature under precise conditions. We can then speculate that quantum horizon-free regular spacetimes may behave as quasiparticles of geometry, as explained below.

\subsection{Quasiparticles of geometry as dark matter?}

The sense of particle-like behavior does not mean that the compact object, that behaves as a gravitational metamaterial, is a particle in the strict sense of particle physics.

So, imaging to model dark matter as \emph{particles of geometry} can just represent an approximation that may deserve further investigation at quantum level.

Nevertheless, in recent developments, it has been shown that coupling scalar fields with curvature and exploring the properties of particle production \cite{Ford:2021syk,Belfiglio:2024xqt} and entanglement generation \cite{Belfiglio:2023moe,Belfiglio:2022cnd} it is possible to infer characteristics of quasiparticles, related to the presence of geometry \cite{Belfiglio:2023rxb}.

These objects have been classified as geometric quasiparticles \cite{Belfiglio:2022qai} and represent an alternative to solitonic contribution in cosmology \cite{Kirillov:2024pcs,Kibble:1976sj,Vilenkin:1984ib}, to inflaton fragmentation  \cite{Gleiser:1993pt,Amin:2010dc} and other objects of interest, such as relics \cite{Dvorkin:2022jyg} or geons \cite{Wheeler:1955zz,Aoki:2017ixz}.

To match our scenario, based on particle-like behavior, with quantum effects and, in particular, to conclude that a gravitational metamaterial is also a quasiparticle of geometry, one expects a collective motion of those objects whose quantum propagator is modified by the presence of geometry and, moreover, under the requirement of stability.

A given free propagator for particles, \textit{i.e.}, the Feynman propagator, may be modified according to the presence of a regular metric, plus a given lifetime associated with an imaginary energy term, following the prescription of field theories.

Hence, to extend our picture to quasiparticles, we may add to the previous list of five requirements, the additional  criteria:
\begin{itemize}
    \item[-] We invoke a modified Feynman propagator under the form
    \begin{equation}\label{propagatore1}
        \Delta\sim \frac{i}{p_\mu p^\mu -m^2_{eff}},
    \end{equation}
    with $p^\mu$ the four momentum and $
        m_{eff}^2\equiv m_b^2+\Sigma(R,\theta)$,    where $\Sigma$ is the contribution due to the possible interaction between the curvature, $R$, and an external fields (scalar, vector, etc., here indicated generically by $\theta$) that contributes with an interacting term to the Lagrangian and modifies the bare mass, $m_b$, associated with the object.
    \item[-]  The stability of such particles can be accounted by the imaginary energy term, modifying Eq. \eqref{propagatore1}, having,
    \begin{equation}
        \Delta\sim \frac{i}{p_\mu p^\mu -m^2_{eff}+i \Gamma_p},
    \end{equation}
that contributes to the lifetime of a compact object having mass $m_{eff}$.

The quasiparticle lifetime therefore demands
\begin{equation}
    t_{{\rm life}}=(2\Gamma_p)^{-1}\leq \frac{1}{H_0},
\end{equation}
where to guarantee stability we fixed $(2\Gamma_p)^{-1}$ to be much smaller than the cosmic scales, here in terms of $H_0$, \textit{i.e.}, the Hubble constant. This scheme provides a full propagator, $\Delta$, that extends Eq. \eqref{propagatore1}, under the form,
\begin{equation}\label{propagatore2}
    \Delta \sim \frac{iZ_p}{p_\mu p^\mu-m_b^2-\Sigma+i H_0/2}+\ldots,
\end{equation}
that contains further contributions from short-lived multiparticle states which can be scattered out of the vacuum. There, moreover, $Z_p$ explains how likely it is for a particle with three-momentum ${\bf p}$ to exist stably without being destroyed by the interactions.
\item[-] The existence of gravitational metamaterials is associated with external fields, $\theta$, nonminimally coupled with geometry, that act to produce regular, and possibly horizon-free, metrics\footnote{See Ref. \cite{Bronnikov:2018vbs} for a detailed discussion on how to obtain regular solutions from additional (scalar) fields in the Hilbert-Einstein action.} that fulfill the five conditions described above and generate, moreover,  quasiparticle behavior, through the propagator \eqref{propagatore2}.
\end{itemize}

Hence, to permit a semiclassical nature of quasiparticles from the above classical solutions, one needs a source generating objects that behave as gravitational metamaterials and quasiparticles, at the same time.

As a possible naive example allowing to obtain the above scheme is likely offered by scalar or fermion fields nonminimally coupled with gravity or by nonminimal dilaton fields, etc. \cite{Belfiglio:2025chv}.

Focusing on scalar fields, one may recall that geometric quasiparticles necessitate to describe dark matter, having possibly that the perturbation sound speed, $c_s$, vanishes.

If this happens the Jeans length,

\begin{equation}
\lambda_J=c_s\sqrt{\frac{\pi}{\rho}},
\end{equation}
is zero, \emph{enabling dark matter to cluster at all scales}.

However, a standard scalar field sector behaves as stiff matter, as in the quintessence framework, thus having $c_s=1$ \cite{DeDeo:2003te}.

Conversely, the simplest Lagrangian, $\mathcal L$, that permits to write the expression in Eq. \eqref{propagatore2}, in terms of a scalar field, $\varphi$, that generates the gravitational metamaterial, can be represented by a K-essence scenario \cite{Armendariz-Picon:2000ulo} or by a more general \emph{matter with pressure Lagrangian} \cite{DAgostino:2022fcx,Lim:2010yk}, reading \cite{Luongo:2018lgy}
\begin{align}\label{eq:1}
\mathcal{L} &= K\left(X,\varphi\right) +\zeta\, Y\left[X,\nu\left(\varphi\right)\right]-V^{eff}\left(\varphi,R\right)\,,
\end{align}
where $K(X,\varphi)$ is a generalized kinetic term, with $X \equiv\frac12 g^{\alpha\beta}\partial_\alpha \varphi \partial_\beta\varphi$,  $V^{eff}(\varphi,R)$ is the potential that drives the dynamics, whereas $\zeta$ is a  Lagrange multiplier, constructed to enforce the total energy constraint of the universe, having that the function $\nu(\varphi)$ governs the specific inertial mass of the scalar field itself \cite{Belfiglio:2024swy}.

This approach, named \emph{quasi-quintessence}, seems more general than quintessence \cite{Luongo:2023aaq}. For one free scalar field, it is possible to set an effective potential made up of a nonminimal coupling between curvature and $\varphi$ as

\begin{equation}\label{massariscalata}
    V^{eff}\equiv \frac{1}{2}m_b^2\varphi^2+\xi R\varphi^2\equiv \frac{1}{2}m^2(R)\varphi^2,
\end{equation}
having moreover $m^2(R)\equiv m_{eff}^2$, guaranteeing the propagator \eqref{propagatore2} to be satisfied and modifying \emph{de facto} the bare mass, $m_b$. The potential has been written, ensuring the simplest scalar field potential of harmonic oscillator and the most practical nonminimal coupling between curvature and $\varphi$ by virtue of a Yukawa-like potential term, $\propto R\varphi^2$. The fact that $m^2(R)\equiv m_{eff}^2$ depends on choosing one single field, $\varphi$.

In this framework, we have the density and pressure under the form,
\begin{align}
\label{eq:no11bis}
\rho =\,&2X \mathcal{L}_{,X} +  V^{eff}(\varphi,R)\,,\\
\label{eq:no12bis}
P =\, &K - V^{eff}(\varphi,R)\,.
\end{align}
Moreover, if the generalized kinetic term does not depend on $\varphi$, it is possible to show that the Noether currents are automatically conserved and the Helmotz energy is positive definite; see e.g. Ref. \cite{Luongo:2018lgy} for additional details. Ensuring the above recipe on $K$ does not imply that $X$ is constant and, so, one easily finds
\begin{subequations}
    \begin{align}
        c_s^2&=\frac{\partial P}{\partial X}\left(\frac{\partial \rho}{\partial X}\right)^{-1}=0,
    \end{align}
\end{subequations}
in which we used the fact that $K_{,X}=0$, as found in Ref. \cite{2010PhRvD..81d3520G}. This property shows that the \emph{effective fluid}, associated with the Lagrangian, emulates matter and appears quite different from quintessence   \cite{Luongo:2023jnb,Luongo:2014nld}.

To create an emergent excitation with collective structure that acts as a quasiparticle, we need $1)$ \emph{more than one excitation}, \textit{i.e.}, more than one field, $\varphi$; 2) \emph{an interaction among all the fields involved}. It is always possible to diagonalize fields of same type, so we can imagine to couple the above lagrangian with another  one and, at the same time, we require that the collective excitations appear optically invisible, analogously to gravitational metamaterials. To do so, we proceed through three steps, below reported.

\begin{itemize}
    \item[a)] We consider a coupling with gravity and fields, having a full Lagrangian, $\mathcal L_{{\rm tot}}$, given by,
\begin{equation}\label{lagrangianatotale}
        \mathcal L_{{\rm tot}}=R+\sum_i \mathcal L_i(X,\varphi_i,\zeta,R)+\mathcal L^\prime(\mathcal F,R,\phi,\varphi,\ldots),
    \end{equation}
    where we added the Hilbert-Einstein Lagrangian, plus $\mathcal L_i$ that represent the $i$-th quasiquintessence sectors, associated with matter and depending on $\varphi$, while $\mathcal L^\prime$ depending on whichever additional sector,  such as electromagnetic $\mathcal F\equiv-\frac{1}{4}F^{\mu\nu}F_{\mu\nu}$, $\phi$  spectator scalar field, and so on.
    Hence, the quasiparticle behavior can be obtained assuming \emph{at least} the interaction between two quasiquintessence fields, say $\varphi_1$ and $\varphi_2$, with a bilinear interacting term and a nonminimal coupling with the Ricci scalar, $R$, namely
    \begin{equation}\label{compactlag}
        \mathcal L_{{\rm tot}}=\sum_{i=1}^2\mathcal L_i(X,\varphi_i,\zeta,R)-g\varphi_1\varphi_2,
    \end{equation}
    which can be significantly simplified, assuming moreover
\begin{subequations}\label{condizionicampi}
\begin{align}
         &V_1^{eff}(\varphi_1)=\frac{1}{2}m_{b,1}^2\varphi_1^2+\xi_1\varphi_1^2R\,,\\
        &V_2^{eff}(\varphi_2)=0\,,\\
        &\xi_1\ll1,
    \end{align}
\end{subequations}
    while having the unspecified coupling constant, $g$, dictating the underlying interaction strength between the two quasiquintessence fields. The conditions in Eqs. \eqref{condizionicampi} are among the simplest ones. They certify that one field is massive, $\varphi_1$ and interacts with gravity, whereas the sound speeds remain zero, guaranteeing that the net fluid behaves similarly to dust.
    \item[b)] Proceeding with the mass mixing, we can diagonalize the mass matrix, having the effective mass provided by:
\begin{equation}\label{matricedimassa}
{\bf M}^2=\begin{pmatrix}
m_1^2(R) & g \\
g & 0
\end{pmatrix},
\end{equation}
where $m_1(R)$ is found in Eq. \eqref{massariscalata}, for the field $\varphi_1$. Note that assuming $m_1^2(R)$ is as much valid as $R$ is roughly constant. This may happen in de Sitter-like configurations, such as inflationary stages, or in regimes of static solutions and weak field. The latter case has also been the case considered in modeling gravitational metamaterials, above.

Accordingly, from Eq. \eqref{compactlag}, we infer the Lagrangian under the form
\begin{equation}
\mathcal L_{{\rm tot}}=K_1+K_2-\frac{1}{2}\begin{pmatrix}
\varphi_1 & \varphi_2
\end{pmatrix}\begin{pmatrix}
m_1^2(R) & g \\
g & 0
\end{pmatrix}\begin{pmatrix}
\varphi_1  \\
\varphi_2
\end{pmatrix},
\end{equation}
yielding new modes of effective mass, written in terms of the bare mass of $\varphi_1$ and of the coupling constants, $g$ and $\xi_1$.

Indeed, assuming two new field excitations, $\Phi_1$ and $\Phi_2$ and diagonalizing Eq. \eqref{matricedimassa}, we obtain two effective masses, $m_{eff}^{+}$ and $m_{eff}^{-}$,
\begin{equation}
    m_{eff}^{2,\pm}=\frac{1}{2}\left(
m_1^2(R)\pm\sqrt{m_1^4(R)+4g^2}\right),
\end{equation}
corresponding to the boson mixing provided by the new fields, $\Phi_1$ and $\Phi_2$:
\begin{eqnarray}\label{mixing}
\Phi_1&=\varphi_1\cos\vartheta+\varphi_2\sin\vartheta,\\
\Phi_2&=\varphi_2\cos\vartheta-\varphi_1\sin\vartheta,
\end{eqnarray}
where the mixing angle turns out to be
\begin{eqnarray}
    \vartheta=\frac{1}{2}\tan^{-1}\left(\frac{2g}{m_{b,1}^2+\xi_1 R}\right).
\end{eqnarray}

Nevertheless, we notice that in case of $m_{b,1}\rightarrow0$, the $m_{eff}^{\pm}$ appear purely geometric dependent, giving rise to \emph{geometric quasiparticles}, whose origins come from the couplings with strength constant, $g$, providing,
\begin{subequations}\label{Rdominante}
\begin{align}
    \vartheta&\simeq \frac{g}{\xi_1R},\\
    m_{eff}^{2,+}&\simeq\xi_1 R-m_{eff}^{2,-}\\
    m_{eff}^{2,-}&\simeq-\frac{g^2}{\xi_1 R},
\end{align}
\end{subequations}
valid as much as $R$ is roughly constant and larger enough than $g$, \textit{i.e.}, suggesting that the mixing angle decreases as $R$ increases, letting curvature generates the quasiparticle mass, suppressing the mixing in Eqs. \eqref{mixing}. Notice that $R$ can be even small, for example in weak field regime, but much larger than $g$, guaranteeing the above approximations. Interestingly, if $R$ is very large, then $m_{eff}^{-}\simeq 0$ and $m_{eff}^{+}\simeq \xi_1 R$.

This mechanism seems to produce prototypes of \emph{geometric quasiparticles} that extends the one field approach \cite{Belfiglio:2024swy} and \emph{appears the source to characterize potential spacetimes that fulfill the five conditions to have gravitational metamaterial optical behavior}.

\item[c)] Once the collective motion defined by the fields $\Phi_1$ and $\Phi_2$ is accounted, the physical properties of a geometric quasiparticle can be found when the effects of curvature are dominant. This corresponds to Eq. \eqref{Rdominante}, above.

In so doing, the quasiparticle states are ``dressed", namely a particle is dressed with interaction between the scalar sector and gravity, as in Eq. \eqref{compactlag}, and provides possible quantum numbers, such as neutral charge, spin inside the interval $[0,2]$, and so on. The so-constructed quasiparticles arise from collective modes and can be imagined as the source for the no-horizon regular solutions that we are searching.

Thus, using Eq. \eqref{metricagenerica}, one can find the regular solution associated with the Lagrangian in Eq. \eqref{lagrangianatotale}, imposing the five conditions to have a particle-like behavior. Hence, supposing regularity and asymptotic flatness, one can work on imposing finite pressure and density, presuming invisibility by ensuring the fifth condition on having $n_O<0$.

The overall treatment, therefore, can explain not only the elusive nature of dark matter as particle-like configurations that behave as geometric quasiparticles, but also the fact that dark matter is invisible and not directly detectable.

Accordingly, classes of spacetimes can be computed from the Lagrangian of coupled scalar fields, above reported. Precisely, a given class of spacetimes originated from Eq. \eqref{lagrangianatotale} goes beyond of the purposes of this work, where specifically we focused on how to produce them. Finding it out  will be object of future efforts, having as target to determine either black hole solutions to regularize or to find regular spacetimes that, in both case, might exhibit no horizons, showing in the meanwhile a negative refraction.

\end{itemize}

\onecolumngrid

\begin{table}[!htb]
\resizebox{\columnwidth}{!}{%
\begin{tabular}{|c|cc|cc|}
\hline\hline
\multirow{2}{*}{Spacetime behavior} & \multicolumn{2}{c|}{Kind of interaction with light} & \multicolumn{2}{c|}{Possible particle behavior} \\ \cline{2-5}
 & \multicolumn{1}{c|}{$n_O$} & Deflection & \multicolumn{1}{c|}{As dark matter} & Optically \\ \hline\hline
Material & \multicolumn{1}{c|}{$>0$} & $\alpha<0, \sigma^\prime=-1-\frac{r}{n_O}\frac{dn_O}{dr}$ & \multicolumn{1}{c|}{Particles \cite{Holzhey:1991bx} or primordial black holes \cite{Villanueva-Domingo:2021spv}} & Dark $\Longrightarrow$ invisible\\ \hline
Metamaterial & \multicolumn{1}{c|}{$<0$} & $\alpha>0, \sigma^\prime=-1-\frac{r}{n_O}\frac{dn_O}{dr}$ & \multicolumn{1}{c|}{Geometric Particle-like or quasiparticles \cite{Belfiglio:2022cnd,Belfiglio:2023moe,Belfiglio:2024swy,Belfiglio:2024xqt}} & Invisible $\Longrightarrow$ Light bent outward\\ \hline
\hline
\end{tabular}%
}
\caption{Summary of our results. Here, material and metamaterial refer to the spacetime behaviors. Gravitational metamaterials as dark matter suggests a geometric contribution to it, within the context of general relativity, quite differently from previous geometric dark matter approaches that, instead, predict geometric contributions extending Einstein's theory \cite{Demir:2020brg}.}
\label{tab: main points}
\end{table}

\twocolumngrid

Last but not least, a final summary of our findings is reported in Table \ref{tab: main points}, resuming the main points, open challenges, and properties of the overall treatment adopted throughout this manuscript.

\section{Final outlooks and perspectives}\label{sezione9}

In this work, we discussed spacetime  optical properties investigating the refractive index, the light deflection and the physics of those objects that may exhibit \emph{an exotic negative refractive index}.

To do so, under the hypothesis that a given spacetime behaves as a medium, we studied two strategies to determine the refractive properties. The first method directly involves the light propagation in the medium, while the second indirectly finds the refractive index, postulating an optical metric.

We limited our analyses in spherical coordinates and obtained an expression for the refractive indexes for both the methods. We thus showed that the two treatments are essentially equivalent and discussed the physical consequences of a \emph{gravitational refractive index}, $n_O$, that can be either positive or negative, depending on the sign of the lapse function.

To analyze the implications of this approach, we examined the cases of positive and negative refractions when $n_O$ is  homogeneous, inhomogeneous, and anisotropic. These scenarios corresponded to a constant or variable $n_O$, while the latter implies that the $n_O$ magnitude also depends on the three-sphere volume.

To better explore the characteristics of a negative refractive index, we related it to the MS mass first and, then, to the redshift.

We emphasized that the increase of the MS mass lets $n_O$ increase, conversely, the decrease of MS mass implies a decrease of $n_O$. Different behaviors are thus investigated for the various singular and regular spacetime candidates.

Afterwards, to generalize our findings, we considered, for weak gravitational fields, how to obtain negative $n_O$. This appears true even for small deflection angles, \emph{indicating that small light deflection can even be recovered in a negative refractive index domain}. The results are tested in generic spacetimes and compared with the Schwarzschild metric.

Accordingly, we considered whether it might be possible to incorporate electromagnetic fields to construct $n_O$ in curved spacetimes. Hence, by analyzing Maxwell's equations in curved spacetimes, we observed how to find the same outcomes, derived purely from general relativity, using electric and magnetic fields. We thus noticed an equivalence between the two frameworks, that holds when the deflection angles are small.

Once discussed generic optical properties, we built up $n_O$ expressions for black holes, regular black holes, naked singularities and solutions with no horizons.

Due to the fact that at $r=0$, the regular black holes here considered, namely the Bardeen, Hayward, Dymnikova and Fang-Wang spacetimes, provide a de Sitter core, then their optical behaviors on $n_O$ appeared much smoother than Schwarzschild. Generally, the increase of $n_O$ is found for all the solutions, \textit{i.e.}, singular or regular, while approaching the horizons, albeit the values of $n_O$ are smoother for regular black hole, being generally smaller than black hole ones.

Analogously, solutions with no horizons appeared extremely smooth across the interval $r\in[0,\infty[$ and, therefore, an eventual observation of optical properties of spacetime can be used to distinguish the three configurations, \emph{trying to use refractive indexes to distinguish black holes from generic mimickers}.

Quite remarkably, we demonstrated that the signs of $n_O$ for regular black holes remain unchanged both inside and beyond the horizons, whereas for standard black holes, say the Schwarzschild, Reissner-Nordstr$\ddot {\rm o}$m, and Schwarzschild-de Sitter solutions, the signs change within the horizons. This appears contradictory, since it violates the expectations for which \emph{inside the horizon}, one needs a change of sign of $n_O$, that passes from positive to be negative and vice versa. This may show some incompleteness of regular black holes, in analogy to recent findings on the repulsive effects of such solutions; see e.g. \cite{repulsivegravity}. At the same time, this result implies that the internal structure of a regular black hole can, in principle, slightly differ from that of standard black holes. Further, it may shed light on exploring quantum  solutions, unveiling the interior structure of regular spacetime media. In this respect, regular black holes are expected to violate the strong energy condition, which could, in principle, indicate the $n_O$ behavior within the horizon.

Toward no-horizon regular spacetimes,  we worked out an AdS metric and a quasi-AdS solution, generally showing that to be distinguished from black holes one has to get closer to the Schwarzschild horizon to check whether $n_O$ increases significantly or appears quite smoothly evolving, as it occurs for naked singularities. At very large distances, instead, there is no chance to distinguish singular, regular or naked solutions.

Interestingly, for naked singularities and, more generally, for no-horizon solutions, however, the refractive index can even acquire values smaller than unity, namely the refractive index may show \emph{superluminal behaviors}. We showed that this occurrence, in the case of AdS and quasi-AdS solutions, may be attributed to the presence of a de Sitter phase. The same, in fact, does not occur for singular or regular solutions. Among other prototypes of naked singularities, we also employed a negative mass counterpart of the Schwarzschild solution and the Reisser-Nordstr{\rm $\ddot o$}m metric with $Q>M$. The corresponding results have thus been compared among them, indicating a change of concavity among de Sitter-like solutions with respect to the above two naked singularities.

Afterwards, we investigated the Snell's law, showing how to obtain the variation of the refractive angles with respect to the azimuthal angle. We computed the variation of $\sigma$ for all the involved spacetimes and showed that the deflection does not change under the switch $n_O\rightarrow-n_O$. We split our analysis for weak and strong fields. In particular, weak gravitational fields have  been analyzed by considering the Sun and Earth configuration, finding a deflection angle perfectly compatible with lensing predictions. Conversely, as strong gravitational sources, we considered neutron stars and white dwarfs. There, we predicted the deflections to lie in larger intervals than the Sun. Physical consequences of these results are thus explored and critically interpreted in view of future observations. Further, we reported theoretical bounds on the de Sitter phases, predicted by de Sitter-like metrics and by the Hayward spacetime, as well as bounds on the electric and topological charges and limits on the free parameters of Dymnikova and Fang-Wang spacetimes.

In our treatment, we thus introduced the concept of \emph{gravitational metamaterial behavior} or briefly of \emph{gravitational metamaterials}, associated with negative refraction induced by given spacetimes. In particular, we presented an analogy between spacetime materials, showing positive refraction, and metamaterials, exhibiting negative refractive indexes, instead. In this respect and, following the modern results on metamaterials in laboratory, we proposed that our  \emph{gravitational metamaterials, may be invisible, as consequence of having $n_O<0$}.

Further, as a first example of gravitational metamaterials in flat space, we also tried to obtain $n_O<0$ for the Rindler space.

There, we encountered a possible inconsistency in having a negative behavior of $n_O$, studying the Unruh temperature. Accordingly, we ended up with the fact that negative refractive indexes do not alter the temperature, as it does not depend on the sign of acceleration, but rather on its modulus.

Hence, starting from the fact that the sign of lapse function can determine the refractive sign even in flat spaces, we remarked that \emph{the role of curved space is to modify the inhomogeneous and anisotropic parts of $n_O$}.

To this end, we finally furnished a physical interpretation to objects which exhibit negative refractive indexes. Premising that
black hole themselves can show such properties, we here proposed that regular metrics, with no horizons and negative refractive indexes can behave as particles, or better they can appear as particle-like configurations, satisfying five basic demands that we enumerated to guarantee such a property.

The particle-like behavior is thus discussed for a conformal version of spacetime that reproduces the simplest homogeneous case in which $n_O$ is constant and negative. Then, we worked out the simplest inhomogeneous case of negative refractive medium adopting a Newtonian metric and imposing the aforementioned conditions. Last but not least, we remarked that a regularization mechanism, proposed by Simpson and Visser can help in regularizing spacetimes, in order to obtain regular media with negative optical properties.

Since these compact objects are thought to be invisible as due to the fact that $n_O<0$, we ensured pressures to be small enough and to limit the solutions to be less relativistic, namely to mime the dark matter constituent.

To this end, we speculated about the existence of geometric dark matter  provided by gravitational metamaterials. Recalling, then, the three toy models here considered, we also showed their advantages and disadvantages in the context of spiral galaxies, computing the kinematic contribution to the radial velocity for the inhomogeneous model.

The approach appeared promising to increase velocity without introducing extra dark matter particles, that extend the Standard Model of particle physics. Conversely, it relies on solutions to Einstein's field equations, without generalizing the background theory itself. Last but not least, those contributions also explained the fact that dark matter does not interact electromagnetically, appearing invisible as consequence of $n_O<0$.

As a further speculation, we wonder whether such particle-like configurations can exhibit a quantum nature. Hence, we seek which quantum source can generate these spacetime media with exotic optical properties, preserving regularity.

Hence, we suggested the concept of \emph{quasiparticles of geometry}, invoking a collective motion and the modification of the Feynman propagator, as consequence of interacting matter-like fields, nonminimally coupled with gravity.

Limiting to two matter fields, interacting between them with a bilinear term, in a Yukawa-like potential in which one field only interacts with geometry, we suggested that such a Lagrangian can reproduce regular black hole solutions, plus additional information on the sound speed and effective masses of constituents. Indeed, to guarantee that our scalar fields mime matter, we suggested that a Lagrange multiplier is involved, modifying the pressure's equation of state and enabling the sound speed of perturbations to identically vanish.

Accordingly, the quantum propagator has been modified switching the bare mass to two effective new values, describing, moreover, the underlying particle lifetimes, bounded to the age of the universe, \emph{de facto} being stable. The corresponding new mass modes, associated with these quasiparticle states, have thus been computed. Their physical properties have been discussed in terms of a roughly constant curvature and of the coupling constant between the two scalar fields.

Future research will investigate how the refractive index evolves in collapsing spacetimes and across transitions between two or more matched metrics. In this context, we aim to determine whether measurable effects could be observed during the collapse of astrophysical objects. Notably, we will also explore the role of the refractive index in cosmological settings, with a particular focus on potential connections between our approach and the Gordon metric \cite{Gordon:1923qva}. Toward the fact that such objects, or better the gravitational metamaterials, may candidate as a fraction of dark matter, we will explore in more detail the particle-like behavior versus the possibility to produce quasiparticles of geometry at primordial times, investigating how from the modified propagator, shown before, it would be possible to infer the dynamical and production rate properties of geometric quasiparticles. This would shed light on the quantum nature of dark matter constituents with the ambitious target to characterize their final properties.

\section*{Acknowledgments}
The author expresses his warm gratitude to Maryam Azizinia for insightful discussion on the role of  metamaterials in solid state physics and on their applications in  material science. He is also thankful to Fabio Marchesoni for interesting debate about the optical properties of metamaterials and their modern interpretation. Sincere acknowledgments to   Roberto Giambò, Marco Muccino, Daniele Malafarina and Hernando Quevedo toward the use refractive index in gravitational domains and suggestions on the numerical parts. This paper is supported by the  Fondazione  ICSC, Spoke 3 Astrophysics and Cosmos Observations. National Recovery and Resilience Plan (Piano Nazionale di Ripresa e Resilienza, PNRR) Project ID $CN00000013$ ``Italian Research Center on  High-Performance Computing, Big Data and Quantum Computing" funded by MUR Missione 4 Componente 2 Investimento 1.4: Potenziamento strutture di ricerca e creazione di ``campioni nazionali di R\&S (M4C2-19)" - Next Generation EU (NGEU).

\bibliographystyle{unsrt}

\end{document}